\documentclass[notitlepage,a4paper,10.5pt]{article}

\usepackage[left=2.5cm,right=2.5cm,top=3cm,bottom=3cm]{geometry}

\usepackage{color}
\usepackage{cite}          
\usepackage{textcomp}
\usepackage{protosem}

\usepackage[all]{xy}
\usepackage{graphicx}      
\usepackage{units}
\usepackage{amsmath}
\usepackage{amsfonts}
\usepackage{amssymb}
\usepackage{mathrsfs}
\usepackage{amsthm}
\usepackage{float}
\usepackage{authblk}
\usepackage{enumitem}
\usepackage{xcolor}
\usepackage{appendix}
\usepackage{booktabs}
\usepackage{array}
\usepackage{colortbl}
\usepackage{tabularx}
\usepackage{threeparttable}
\usepackage{indentfirst}

\usepackage[colorlinks=true,
            linkcolor=blue,
            citecolor=blue,
            urlcolor=blue]{hyperref}

\newcommand{\mc}{\mathcal}

\newcommand{\bol}{\boldsymbol}

\newcommand{\abs}[1]{\left\lvert{#1}\right\rvert}
\newcommand{\w}{\wedge}
\newcommand{\lr}[1]{\left({#1}\right)}
\newcommand{\lrs}[1]{\left[{#1}\right]}

\newcommand{\mf}{\mathfrak}
\newcommand{\p}{\partial}

\newtheorem{thm}{\textit{Theorem}}[section]

\theoremstyle{definition}
\newtheorem{mydef}[thm]{Definition}

\theoremstyle{remark}
\newtheorem{remark}[thm]{\textbf{Remark}}

\newtheorem{prop}[thm]{\textbf{Proposition}}
\newtheorem{example}[thm]{\textit{Example}}
\newtheorem{lemma}[thm]{\textbf{Lemma}}
\newtheorem{q}[thm]{\textbf{Question}}

\newcommand{\eq}[1]{\begin{equation}\begin{split}{#1}\end{split}\end{equation}}
\newcommand{\sys}[2]{\begin{subequations}\begin{align}{#1}\end{align}\label{#2}\end{subequations}}

\begin{document}

\title{Topological Invariants in Higher-Dimensional Magnetohydrodynamics}
\author{Naoki Sato\thanks{National Institute for Fusion Science,  322-6 Oroshi-cho Toki-city, Gifu 509-5292, Japan,\ Email: \href{sato.naoki@nifs.ac.jp}{sato.naoki@nifs.ac.jp}
}  
{}\thanks{Graduate School of Frontier Sciences, The University of Tokyo,  
Kashiwa, Chiba 277-8561, Japan}
\qquad Ken Abe \thanks{Department of Mathematics, Graduate School of Science,  Osaka Metropolitan University 3-3-138, Sugimoto, Sumiyoshi-ku, Osaka 558-8585, Japan, Email: \href{kabe@omu.ac.jp}{kabe@omu.ac.jp}} \qquad 
Michio Yamada \thanks{Research Institute for Mathematical Sciences, Kyoto University, Kyoto 606-8502, Japan,\ Email: \href{yamada@kurims.kyoto-u.ac.jp}{yamada@kurims.kyoto-u.ac.jp}}}

\date{\today}
\setcounter{Maxaffil}{0}
\renewcommand\Affilfont{\itshape\small}

    \maketitle
    
\begin{abstract}
It is well known that the three-dimensional ideal magnetohydrodynamics (MHD) equations possess three magnetic invariants: (M) magnetic helicity, (C) cross helicity, and (P) the mean-square magnetic potential, in addition to the fundamental invariants of fluid motion. 
In this paper we construct higher-dimensional generalizations of these invariants for ideal MHD. 
Specifically, we identify generalized magnetic helicity and generalized cross helicity in all odd spatial dimensions $n=2m+1$, and families of invariants given by integrals of arbitrary functions of the scalar density $B^m/\nu$ of the magnetic field $2$-form $B$, where $B^m$ denotes its $m$-fold wedge product and $\nu$ the fluid-density top form, in all even spatial dimensions $n=2m$. 
We further establish the existence of invariants for symmetric solutions in arbitrary dimensions, generalizing the mean-square magnetic potential and showing that this invariant arises from symmetry rather than from even dimensionality, in contrast to the enstrophy invariant of the two-dimensional Euler equations.
\end{abstract}

\tableofcontents

\section{Introduction} 

\subsection{Magnetic helicity and MHS equilibria} 

 In the absence of dissipation, ideal magnetohydrodynamics (MHD) plasmas \cite[pp.107--108]{Fitz}, \cite[p.7]{Freid} evolving in three-dimensional Euclidean space exhibit a number of conservation laws.  
Besides the fundamental invariants related to fluid motions, such as total mass, total energy, and total momentum, topological invariants play a central role in constraining the admissible dynamics and characterizing the global structure of the flow. For instance, the magnetic helicity
\begin{equation}
\mathscr{M} = \int_{\Omega} \boldsymbol{A} \cdot \boldsymbol{B} \, dV, \label{eq: MH}
\end{equation}
defined for the magnetic field \(\boldsymbol{B}\) satisfying $\nabla \cdot \bol{B}=0$ and $\bol{B}\cdot \bol{n}=0$ on $\partial\Omega$ with its vector potential \(\boldsymbol{A}\) is conserved for smooth solutions to ideal MHD on a bounded and simply-connected domain \(\Omega \subset \mathbb{R}^3\), e.g., \cite[III, 1.C]{ArnoldKh}, where $\bol{n}$ is the unit outward normal on $\partial\Omega$ and $dV$ is a volume form on $\Omega$. The magnetic helicity quantifies the linking of magnetic field lines~\cite{Moffatt69}, \cite{Moffatt21}, and constrains the magnetic field to evolve in a topology-preserving manner.

The significance of such invariants extends beyond the ideal regime.  
When dissipative effects, such as viscosity or resistivity, are introduced, invariants are generally no longer conserved. However, their relative robustness allows them to serve as effective constraints on long-term dynamics, e.g., \cite{Hasegawa85}, \cite{Yoshida02}, \cite{Faraco}. In many cases, steady states arise as minimizers of energy under the constraint of a nearly conserved helicity.  
A notable example is Taylor's relaxation theory~\cite{Taylor74}, \cite{Taylor86}, in which resistive dissipation leads to a relaxed magneto-hydrostatic (MHS) equilibrium state constrained by magnetic helicity conservation, resulting in a force-free (Beltrami) configuration. Identifying the invariants of a fluid theory thus provides crucial insight into the qualitative behavior of the system, both in the ideal limit and in the presence of weak dissipation.

Moreover, the conservation of magnetic helicity is closely connected to the question of flexibility and rigidity of three-dimensional MHS equilibria~\cite{Grad67}, \cite{Grad85}, \cite{CDG21b}. 
Indeed, the idea of magnetic relaxation~\cite{Moffatt85}, \cite{Moffatt21} is to obtain a non-trivial MHS equilibrium as the long-time limit of viscous MHD, starting from an arbitrary initial magnetic field, by exploiting the conservation of magnetic helicity. 
This idea was recently realized in~\cite{CP22} by adopting the Voigt approximation for the three-dimensional viscous MHD on the flat $3$-torus as well as in a bounded domain; see also~\cite{Bhatt}. 
Another important relaxation model on the flat $n$-torus, for $n\geq 2$, has been developed in~\cite{BFV21}; see also~\cite{BKS23}, \cite{JINTAN}. At present, no existence results appear to be known for higher-dimensional MHS equilibria, cf. \cite[II.6]{ArnoldKh}, \cite{KMS23}.

\subsection{The higher-dimensional ideal MHD}

The purpose of this paper is to investigate higher-dimensional MHS equilibria from a viewpoint of topological conservation laws. 
Namely, we derive an intrinsic (coordinate-independent) representation of the inhomogeneous ideal MHD equations by differential forms, starting from the classical formulation of ideal MHD for three-dimensional vector fields. 
More specifically, we consider the ideal MHD equations on a compact and orientable $n$-dimensional Riemannian manifold $\Omega$ for $n \geq 2$, describing the evolution of the unknowns $(\nu,u,B,h)$:
\begin{equation}
\begin{aligned}
&\nu_t+\mf{L}_{\bol{u}}\nu=0,\\
&u_t+\mf{L}_{\bol{u}}u+dh=
-\frac{1}{\rho}\iota_{\bol{J}}B
,\\
&B_t+\mf{L}_{\bol{u}}B=0,\\
&\mu_0\bol{J}=(\delta B)^{\sharp},\\
&dB=0,
\end{aligned}
\label{eq: IMHD}
\end{equation}
where $\nu$, $u$, $B$, and $h$  denote the fluid density ($n$-form), the fluid velocity ($1$-form), the magnetic field ($2$-form), and the pressure function ($0$-form), respectively. The symbol $\mf{L}_{\bol{u}}$ denotes the Lie derivative along a vector field $\bol{u}=u^{\sharp}$ and $\iota_{\bol{J}}$ denotes the inner product with the current field $\bol{J}$. The symbols $d$ and $\delta$ denote the exterior derivative and its adjoint. The function $\rho=*\ \nu$ is the fluid mass density for the Hodge-star operator $*$, and the constant $\mu_0>0$ is the vacuum permeability. The system \eqref{eq: IMHD} describes the inhomogeneous ideal MHD, including the homogeneous (incompressible) case $\rho=\rho_0$ as a particular case. We assume  that the pressure $\tilde{h}=h+\frac{1}{2}|\bol{u}|^{2}$ is barotropic for the non-constant density case, i.e., $\tilde{h}=\tilde{h}(\rho)$. 

The geometric formulation of the ideal fluid equations originates from the seminal work of Arnold~\cite{A66}; see also the monograph~\cite{ArnoldKh} and the recent reviews~\cite{KMS23,DE}. 
For the ideal MHD equations, geometric formulations were developed in the works of Newcomb~\cite{New}, Marsden \emph{et al.}~\cite{MRW}, and Holm \emph{et al.}~\cite{Holm98}; see also~\cite[I.10]{ArnoldKh} and the more recent contributions of Gilbert and Vanneste~\cite{GV,GV2}. 
These geometric approaches have further been extended to related systems such as Hall MHD~\cite{HHS}, extended MHD (XMHD)~\cite{Mo}, and rotating shallow-water MHD~\cite{HHS2}. 
The intrinsic system~\eqref{eq: IMHD} regards the magnetic field as a differential $2$-form in all spatial dimensions, in contrast to the extrinsic generalization 
treating the magnetic field as a divergence-free vector field,  cf.~\cite{MW},\cite{KC}, \cite[I.10, Remark 10.8]{ArnoldKh}.

The system \eqref{eq: IMHD} 
describes MHS equilibria when $u=0$ and the density is constant $\rho=\rho_0$:
\begin{equation}
\begin{aligned}
&dh=
-\frac{1}{\rho_0}\iota_{\bol{J}}B
,\\
&
\mu_0\bol{J}=(\delta B)^{\sharp},\\
&dB=0. 
\end{aligned}
\label{eq: IMHS}
\end{equation}
These intrinsic equilibria share a similar mathematical structure with steady Euler flows for the fluid velocity $1$-form on a manifold \cite{A66}, \cite{ArnoldKh} though the two systems are different for $n\geq   4$; see Table \ref{t: EMHS}.

\begin{table}[h]
\centering
\begin{tabular}{ccc}
\hline
\multicolumn{1}{|c|}{}              & \multicolumn{1}{c|}{\begin{tabular}[c]{@{}c@{}}Steady Euler flows \\ for the fluid velocity\end{tabular}}                                                                 & \multicolumn{1}{c|}{\begin{tabular}[c]{@{}c@{}}MHS equilibria \\ for the magnetic field\end{tabular}}                                                                                                   \\ \hline
\multicolumn{1}{|c|}{Unknowns}                                                                 & \multicolumn{1}{c|}{$u=u_idx^i$, $\Pi$}                                                                                                                                   & \multicolumn{1}{c|}{$B_{ij}=\frac{1}{2}B_{ij}dx^{i}\wedge dx^{j}$, $B_{ij}=-B_{ji}$, $h$}                                                                                                               \\ \hline
\multicolumn{1}{|c|}{\begin{tabular}[c]{@{}c@{}}Equations\\
\end{tabular}} & \multicolumn{1}{c|}{\begin{tabular}[c]{@{}c@{}}$\iota_{\bol{u}}w+d \Pi=0$\\ $\textrm{div}\ \bol{u}=0$\\ $w=d u$\end{tabular}}                                   & \multicolumn{1}{c|}{\begin{tabular}[c]{@{}c@{}}$dh=-\iota_{\bol{J}}B$\\ $dB=0$\\ $\bol{J}=(\delta B)^{\sharp}$\end{tabular}}                                                                            \\ \hline
\multicolumn{1}{|c|}{Expression in $\mathbb{R}^{n}$}                                                     & \multicolumn{1}{c|}{\begin{tabular}[c]{@{}c@{}}$u_{i}(\partial_iu_{j}-\partial_ju_{i})+\partial_j\Pi=0$, $1\leq j\leq n$\\ $\partial_iu_{i}=0$\end{tabular}}              & \multicolumn{1}{c|}{\begin{tabular}[c]{@{}c@{}}$\partial_j h=B_{ij}\partial_k B_{ki}$, $1\leq j\leq n$\\ $\partial_i B_{jk}+\partial_j B_{ki}+\partial_k B_{ij}=0$,\\ $1\leq i,j,k\leq n$\end{tabular}} \\ \hline
\multicolumn{1}{|c|}{$3$D vector form}                                                         & \multicolumn{1}{c|}{\begin{tabular}[c]{@{}c@{}}$\bol{w} \times \bol{u}+\nabla \Pi=0$\\ $\nabla \cdot \bol{u}=0$\\ $\bol{w}=\nabla \times \bol{u}$\end{tabular}} & \multicolumn{1}{c|}{\begin{tabular}[c]{@{}c@{}}$\nabla h=\bol{J}\times \bol{B}$\\ $\nabla \cdot \bol{B}=0$\\ $\bol{J}=\nabla \times \bol{B}$\end{tabular}}                                              \\ \hline
            &                              & \multicolumn{1}{l}{} 
\end{tabular}
\caption{Steady Euler flows and MHS equilibria}\label{t: EMHS}
\end{table}

\subsection{Invariants of the Euler equations and ideal MHD}

Serre~\cite{Ser} and Dezin~\cite{Dez} identified higher-dimensional invariants of the incompressible Euler equations, corresponding to the case $B=0$ and $\rho=\rho_0$ in~\eqref{eq: IMHD}, in odd spatial dimensions $n=2m+1$. 
The even-dimensional invariants are attributed to Tartar~\cite{Ser,Serre18}. 
On manifolds with boundary, these invariants take the form  
\begin{align}
\mathscr{H} &= \int_{\Omega} u \wedge w^{m}, \label{eq: He}\\
\mathscr{E} &= \int_{\Omega} f\!\left(\frac{w^{m}}{dV}\right)dV, \label{eq: En}
\end{align}
for the vorticity $2$-form $w = du$, in odd and even dimensions respectively \cite{OKC}, \cite{KC}, \cite[I.9, Theorem 9.2]{ArnoldKh}. 
In even dimensions, the $2$-form $w$ defines the volume form ($m$-fold wedge product) $w^{m}=w\w...\w w$, and the quantity $\int_{\Omega} w^{m}$ is conserved under the Lie-advection equation $w_t + \mathcal{L}_{\boldsymbol{u}} w = 0$. 
More generally, the integral~\eqref{eq: En} is conserved for arbitrary functions $f$. 
In two dimensions ($m=1$), the invariant~\eqref{eq: En} reduces to the classical generalized enstrophy~\cite{ZY17,IK17}.

For the three-dimensional ideal MHD equations in a bounded and simply connected domain~$\Omega$, the quantities~\eqref{eq: He} and~\eqref{eq: En} are no longer invariants. 
Instead, the following magnetic helicity, cross helicity, and generalized mean-square magnetic potential are conserved:
\begin{align}
\mathscr{M} &= \int_{\Omega} \boldsymbol{A} \cdot \boldsymbol{B}\, dV, \label{eq: MH2}\\
\mathscr{C} &= \int_{\Omega} \boldsymbol{u} \cdot \boldsymbol{B}\, dV, \label{eq: C1}\\
\mathscr{P} &= \int_{\Sigma} f(\Psi)\, dV_{\Sigma}. \label{eq: P1}
\end{align}
The magnetic helicity~\eqref{eq: MH2} and the cross helicity~\eqref{eq: C1} take analogous forms to the fluid helicity~\eqref{eq: He}; their integrands correspond to the volume forms $A \wedge B$ and $u \wedge B$, where $A = \boldsymbol{A}^{\flat}$ and $u = \boldsymbol{u}^{\flat}$. 
In contrast to~\eqref{eq: He} and~\eqref{eq: C1}, the definition of the magnetic helicity~\eqref{eq: MH2} requires the introduction of a vector potential~$\boldsymbol{A}$. 
In particular, when~$\Omega$ has a nontrivial first cohomology group, the definition of~\eqref{eq: MH2} must be made independent of the choice of gauge for~$\boldsymbol{A}$. 
See, e.g.,  \cite{Khesin21} for the definition of magnetic helicity on a three-dimensional manifold with a trivial first cohomology group. 
The cross helicity~\eqref{eq: C1}, on the other hand, measures the linkage between vortex lines and magnetic field lines.


When the magnetic field $\boldsymbol{B}$ is translationally symmetric in $\Omega = \Sigma \times \mathbb{R}$, the quantity~\eqref{eq: P1}, expressed in terms of the flux function $\Psi$ on a bounded and simply connected domain $\Sigma \subset \mathbb{R}^{2}$, is conserved. 
Namely, there exists a unique $\Psi$ such that 
$
\boldsymbol{B} = \nabla \times (\Psi \nabla x^{3}) + B_{3} \nabla x^{3},
$
and $\Psi$ is advected on $\Sigma$; see, for example,~\cite{Moffatt96}. 
In particular,~\eqref{eq: P1} is conserved for the two-dimensional ideal MHD equations. 
For the specific choice $f(t) = t^{2}$, the quantity~\eqref{eq: P1} corresponds to the mean-square magnetic potential~\cite{Bis03}. 

The three-dimensional ideal MHD invariants~\eqref{eq: MH2},~\eqref{eq: C1}, and~\eqref{eq: P1} remain finite for magnetic fields with finite energy and they form a variational structure for MHS equilibria~\eqref{eq: IMHS} in contrast to the Euler invariants~\eqref{eq: He} and~\eqref{eq: En}. This motivates the following question (cf.~\cite[Question~5]{Serre18}):
\begin{q}\label{q: HI}
Are there invariants of the higher-dimensional ideal MHD equations~\eqref{eq: IMHD}?
\end{q}

To the best of the authors' knowledge, no invariants have previously been identified for the higher-dimensional ideal MHD equations with $n \ge 4$.






\subsection{The statement of the main results}

We identify invariants of the higher-dimensional ideal MHD system~\eqref{eq: IMHD}, thereby answering Question~\ref{q: HI}. 
Our first result generalizes the three-dimensional magnetic helicity~\eqref{eq: MH2} and cross helicity~\eqref{eq: C1} to all odd spatial dimensions, and establishes a new invariant in all even dimensions, analogous in form to the generalized enstrophy~\eqref{eq: En}. 
In what follows, we employ the inclusion map $i: \partial\Omega \hookrightarrow \Omega$ and its pullback $i^{*}$.


\begin{thm}[Invariants in odd and even dimensions]\label{t: thm}
Let $\Omega$ be a compact and orientable $n$-dimensional Riemannian manifold with smooth boundary for $n\geq 2$ with trivial first and second cohomology groups. Then, smooth solutions $(\nu,u,B,h)$ to the ideal MHD equations \eqref{eq: IMHD} satisfying the boundary conditons $\bol{u}\cdot \bol{n}=0$ and $i^{*}B=0$ on $\partial\Omega$ admit the following invariants:

\noindent
(i) For $n=2m+1$, 
\begin{align}
\displaystyle \mathscr{C}&=\int_{\Omega}u\w B^m,     \label{eq: C} \\
\displaystyle \mathscr{M}&=\int_{\Omega}A\w B^{m},    \label{eq: M}
\end{align}
where $A$ is a $1$-form satisfying $B=dA$.

\noindent
(ii) For $n=2m$, 
\begin{equation}
\mathscr{W}=
\int_{\Omega}f\left(\frac{B^{m}}{\nu}\right)\nu,   \label{eq: W}
\end{equation}
for arbitrary functions $f$.
\end{thm}

\begin{remark}
The assumption that the first cohomology group vanishes is required for the unique, gauge-independent definition of the magnetic helicity~\eqref{eq: M}, whereas the vanishing of the second cohomology group ensures that the magnetic field is exact, a condition needed for its conservation. The other quantities \eqref{eq: C} and \eqref{eq: W} are conserved on any manifold with a smooth boundary and with non-trivial cohomology groups. The conservation of magnetic helicity \eqref{eq: M} holds also on manifolds with non-trivial first and second cohomology groups for smooth solutions to \eqref{eq: IMHD} for exact initial magnetic $2$-form $B_0=dA_0$ and some $1$-form $A_0$ such that $i^{*}A_0=0$ on $\partial\Omega$ (Theorem \ref{t: thmg}).
\end{remark}

The two-dimensional generalized enstrophy~\eqref{eq: En} of the Euler equations and the generalized mean-square magnetic potential~\eqref{eq: P1} of the ideal MHD equations are invariants of analogous form. 
However, these two invariants differ in nature: the generalized enstrophy~\eqref{eq: En} is an invariant for all even-dimensional solutions, corresponding to~\eqref{eq: W} in ideal MHD, whereas our second result shows that the generalized mean-square magnetic potential~\eqref{eq: P1} is an invariant for symmetric solutions of~\eqref{eq: IMHD} in arbitrary spatial dimensions.


\begin{thm}[Generalized mean-square magnetic potential]\label{t: thm2}
Let $\Sigma$ be a compact and orientable $(n-1)$-dimensional Riemannian manifold with smooth boundary for $n\geq 3$ with trivial first and second cohomology groups. Let $\Omega=\Sigma \times \mathbb{R}$. Let $(\nu,u,B,h)$ be a smooth solution to the ideal MHD equations \eqref{eq: IMHD} satisfying the boundary conditons $\bol{u}\cdot \bol{n}=0$ and $i^{*}B=0$ on $\partial\Omega$. Let $\sigma$ be an $(n-1)$-form such that $\nu=\sigma\wedge dx^{n}$. Assume that the coefficients of $(\nu,u,B,h)$  are independent of the $x^n$-variable. Then, there exists {a $1$-form}  $A=A_{i}\,dx^{i}$ whose components $A_i$ are independent of $x^n$ and {satisfy}  $B=dA$,  and the quantity  
\begin{equation}
\mathscr{P}=
\int_{\Sigma}f\lr{A_n}\sigma,   \label{eq: P}
\end{equation}
is invariant for arbitrary functions $f$.
\end{thm}


The generalized mean-square magnetic potential~\eqref{eq: P} is an invariant for symmetric solutions of the ideal MHD system~\eqref{eq: IMHD}, associated with the $n$-th component of the $1$-form $A = A_{i}\,dx^{i}$, where $\partial_{n}$ denotes the symmetric direction. 
For the Euler equations, the corresponding $1$-form is the fluid velocity $u = u_{i}\,dx^{i}$, and we obtain a similar invariant
\begin{equation}
\mathscr{Q} =
\int_{\Sigma} f(u_{n})\,\sigma. \label{eq: Q}
\end{equation}
The component $u_{n}$ is referred to as the {swirl} in the case of three-dimensional axisymmetric solutions of the Euler equations. 
See~\cite{Moffatt97} for a discussion of axisymmetric solutions of the ideal MHD equations.  

We also establish the conservation of total mass, total energy, and total momentum for the ideal MHD system~\eqref{eq: IMHD} under suitable boundary conditions; see Table~\ref{tab:conservation} and the Appendix.



\begin{table}[h]
    \centering
    \resizebox{1.00\linewidth}{!}{
    \begin{tabular}{>{\bfseries}l >{\centering\arraybackslash}m{6cm}>{\centering\arraybackslash}m{3.5cm}}
        \toprule
        Conservation Laws & Expression & Notes \\ 
        \midrule
        \addlinespace[6pt]
        \rowcolor{gray!20}
        \multicolumn{3}{l}{\rule{0pt}{2.6ex}\textsc{(I) Fundamental Invariants}\rule[-0.8ex]{0pt}{0pt}}\\
        \addlinespace[4pt]
        Total Mass & $\displaystyle N=\int_{\Omega} \nu$  & - \\ 
        Total Energy & $\displaystyle H=\int_{\Omega}\lr{\frac{\abs{\bol{u}}^2}{2}+\mc{U}(\rho)+\frac{\abs{B}^2}{4\mu_0\rho}}\nu$  & $\mc{U}(\rho_0)\geq 0$ \\ 
        Total $i$th Momentum & $\displaystyle P^i=\int_{\Omega}u^i\nu$ & - \\ 
        \midrule
        \addlinespace[6pt]
        \rowcolor{gray!20}
        \multicolumn{3}{l}{\rule{0pt}{2.6ex}\textsc{(II) Invariants related to vorticity}\rule[-0.8ex]{0pt}{0pt}}\\
        \addlinespace[4pt]      
        Generalized Fluid Helicity & $\displaystyle \mathscr{H}=\int_{\Omega}u\w w^m$  &  Odd  $(B=0)$
        \\
     Generalized Enstrophy & $\displaystyle \mathscr{E}=\int_{\Omega}f\lr{\frac{w^{m}}{\nu}}\nu$ & Even $(B=0)$
      \\  
     Analog to Generalized Mean-Square Magnetic Potential & $\displaystyle \mathscr{Q}=\int_{\Sigma}f\lr{u_n}\sigma$ & Under symmetry  $(B=0)$ \\

        \midrule
        \addlinespace[6pt]
        \rowcolor{blue!20}
        \multicolumn{3}{l}{\rule{0pt}{2.6ex}\textsc{(III) Invariants related to magnetic fields}\rule[-0.8ex]{0pt}{0pt}}\\
        \addlinespace[4pt]
        Generalized Cross Helicity  & $\displaystyle \mathscr{C}=\int_{\Omega}u\w B^m$ & Odd \\ 
     Generalized Magnetic Helicity  & $\displaystyle \mathscr{M}=\int_{\Omega}{A}\w B^{m}$ & Odd  \\ 
    Analog to Generalized Enstrophy & $\displaystyle \mathscr{W}=\int_{\Omega}f\lr{\frac{B^{m}}{\nu}}\nu$ & Even  \\  
       Generalized Mean-Square Magnetic Potential  & $\displaystyle \mathscr{P}=\int_{\Sigma}f\lr{A_n}\sigma $ &  Under symmetry     \\ 
        \bottomrule
    \end{tabular}
    }
    \vspace{5pt}
    \caption{The ideal MHD invariants of the system \eqref{eq: IMHD} categorized into (I), (II), and (III) (the main results of this paper)}
    \label{tab:conservation}
\end{table}

We note that Tao \cite{Tao18}, \cite{Tao20} recently introduced the concept of the universality for the incompressible Euler equations, which states that a broad class of dynamical systems can be embedded into the higher-dimensional Euler equations on Riemannian manifolds by choosing a suitable metric, even though solutions are constrained by the invariants \eqref{eq: He}, \eqref{eq: En}, and \eqref{eq: Q}. See Lizaur \cite{Liz}. 

The metric also plays a crucial role in the flexibility and rigidity of three-dimensional MHS equilibria. 
Constantin \emph{et al.}~\cite{CDG21} constructed a nearly quasi-symmetric MHS equilibrium with a small residual force in a toroidal domain by designing a suitable metric that renders a given vector field Killing. 
In contrast, Cardona \emph{et al.}~\cite{CDP} recently demonstrated that MHS equilibria are generically asymmetric, in the sense that for a given MHS equilibrium, most of the metrics adapted to the equilibrium do not admit Killing vector fields, in contrast to the classical conjecture of Grad \cite{Grad67}, \cite{Grad85}, \cite{CDG21b} on the symmetry of MHS equilibria in Euclidean space with flat metric.


\subsection{Organization of this paper} 

In Section 2, we derive the ideal MHD equations on a Riemannian manifold \eqref{eq: IMHD}. In Section 3, we show the conservation of even-dimensional invariants \eqref{eq: En} and \eqref{eq: W}. In Section 4, we show the conservation of odd-dimensional invariants \eqref{eq: He}, \eqref{eq: C}, and  \eqref{eq: M}. In Section 5, we show the conservation of generalized mean-square magnetic potential \eqref{eq: Q}. In Appendix A, we show the conservation of total mass, total energy, and total momentum.

\subsection{Acknowledgments}
The research of N.S. was partially supported by JSPS KAKENHI Grant 
No. 25K07267, No. 22H04936, and No. 24K00615. 
This work was partially supported by the Research Institute for
Mathematical Sciences, an International Joint Usage/Research
Center located in Kyoto University. The research of K.A. was partially supported by the JSPS through the Grant in Aid for Scientific Research (C) 24K06800, MEXT Promotion of Distinctive Joint Research Center Program JPMXP0723833165, and Osaka Metropolitan University Strategic Research Promotion Project (Development of International Research Hubs).

\section{The higher-dimensional ideal MHD}


In this section, we generalize the ideal MHD equations for three-dimensional vector fields in Euclidean space to a formulation in terms of differential forms on an $n$-dimensional compact and orientable Riemannian manifold with boundary, for $n \ge 2$. 
In this generalized setting, the vorticity and magnetic fields evolve according to Lie-advection equations and preserve their three-dimensional properties, expressed by Helmholtz's laws and Alfvén's theorem. 
We revisit the Hodge--Morrey decomposition of differential forms 
and discuss intrinsic MHS equilibria for the magnetic field $2$-form.

\subsection{The three-dimensional ideal MHD for vector fields}

The inhomogeneous ideal MHD equations in a three-dimensional bounded domain $\Omega\subset \mathbb{R}^{3}$ take the form, e.g., \cite[pp.107--108]{Fitz}, \cite[p.7]{Freid}: 
\sys{
&\frac{\p\rho}{\p t}+\nabla\cdot\lr{\rho\bol{u}}=0,\\
&\rho\frac{\p\bol{u}}{\p t}+\rho\bol{u}\cdot\nabla\bol{u}+\nabla P=\bol{J}\times\bol{B},\label{iMHDut}\\
&\frac{\p\bol{B}}{\p t}-\nabla\times\lr{\bol{u}\times\bol{B}}=0,  \label{iMHDBt}\\
&\mu_0\bol{J}=\nabla\times\bol{B}, \label{iMHDJ}\\
&\nabla\cdot\bol{B}=0,
}{iMHD3}
where $\rho\lr{\bol{x},t}$ is the fluid mass density, $\bol{u}\lr{\bol{x},t}=(u^{1}(\bol{x},t),u^{2}(\bol{x},t),u^{3}(\bol{x},t))$ is the fluid velocity, $\bol{B}\lr{\bol{x},t}=(B^{1}(\bol{x},t),B^{2}(\bol{x},t),B^{3}(\bol{x},t))$ is the magnetic field, $P(\bol{x},t)$ is the pressure function and $\mu_0\bol{J}$ is the current density with the vacuum permeability $\mu_0>0$. We assume that $P$ is a barotropic pressure $P=P(\rho)$ if the density $\rho\lr{\bol{x},t}$ is not a constant $\rho\lr{\bol{x},t}\neq \textrm{const.}$ The equations~\eqref{iMHD3} are a system for the vector fields $(\rho, \boldsymbol{u}, \boldsymbol{B}, P)$.

\subsection{The ideal MHD for differential forms}

We generalize the system~\eqref{iMHD3}, defined for the vector fields $(\rho, \boldsymbol{u}, \boldsymbol{B}, P)$, to a system for differential forms $(\nu, u, B, h)$ on a compact and orientable $n$-dimensional Riemannian manifold $\Omega$ with boundary $\partial\Omega$, for $n \ge 2$. 
Let $g_{ij}$ denote the 
Riemannian metric tensor with determinant ${g}$ on $\Omega$. In coordinates, the volume form is $dV=\sqrt{g}\,dx^1\w ...\w dx^n$. We set
\begin{align*}
&\nu=\rho\sqrt{g}\,dx^1\w ...\w dx^n,\\
&u=u_idx^i,\\
&B=\frac{1}{2}B_{ij}dx^i\wedge dx^j,\ B_{ij}=-B_{ji}.
\end{align*}
with $u_i=g_{ij}u^j$. For $n=3$, we also  define $B_{12}=B^{3}$, $B_{23}=B^{1}$, $B_{31}=B^{2}$. We use the symbols $u=\bol{u}^{\flat}$ and $\bol{u}=u^{\sharp}$. For $\rho\lr{\bol{x},t}= \rho_0>0$, we set $h=P/\rho_0-|\bol{u}|^{2}/2$ with $\abs{\bol{u}}^2=u^iu_i$. For $\rho\lr{\bol{x},t}\neq \textrm{const.}$, we set $h$ by ${h}=\tilde{h}-\abs{\bol{u}}^2/2$ by a primitive function $\tilde{h}\lr{\rho}$ of $\rho P'(\rho)$ so that $d\tilde{h}=dP/\rho$.


We denote the interior derivative with a vector field $\bol{u}$ by 
$\iota_{\bol{u}}$ and the Lie derivative by $\mf{L}_{\bol{u}}=\iota_{\bol{u}} d+d\iota_{\bol{u}}$. We use $
\lr{\nabla_{\bol{u}}\bol{u}}^{\flat}
=\mf{L}_{\bol{u}} u-d(|\bol{u}|^{2}/2)$ and $\bol{J}\times \bol{B}=-\iota_{\bol{J}}B$. We denote the Christoffel symbols by $\Gamma^i_{jk}$ and the covariant derivative by $\nabla^k=g^{k\ell}\nabla_{\ell}$, which acts on a covariant $2$-tensor $T_{ij}$ according to $\nabla_\ell T_{ij}=\p_\ell T_{ij}-\Gamma^k_{\ell i}T_{kj}-\Gamma^{k}_{\ell j}T_{ik}$. We express the components of the bivector field $B^{\sharp}$ corresponding to $B$ by $B^{ij}=g^{ik}g^{j\ell}B_{k\ell}$ and denote the tensorial divergence of $B^{\sharp}$ by ${\rm div}B^{\sharp}=\lr{\nabla_j B^{ji}}\p_i$. We express $\mu_0\bol{J}=\nabla \times \bol{B}$ and $\nabla \cdot \bol{B}=0$ by $\mu_0\bol{J}=-{\rm div}B^{\sharp}$ and $dB=0$. We also express $\bol{u}\times \bol{B}$ and $\nabla \times (\bol{u}\times \bol{B})$ by $\iota_{\bol{u}}B$ and $\mf{L}_{\bol{u}} B$. By using the codifferential $\delta = (-1)^{3n+1} *\ d\ *$ acting on $2$-forms, $-{\rm div}B^{\sharp}=\lr{\delta B}^{\sharp}$.

We assume compactness and orientability for the Riemannian manifold $\Omega$ to integrate differential forms. The following intrinsic generalization of \eqref{iMHD} can be defined without 
compactness and orientability (by replacing $\lr{\delta B}^{\sharp}$ with $-{\rm div}B^{\sharp}$).

\begin{mydef}[Ideal MHD] Let  
$\Omega$ be an $n$-dimensional compact and orientable Riemannian manifold with boundary $\p\Omega$ for $n\geq 2$. Let $(\nu, u, B, h)$ be a collection of differential forms. Let $\rho=*\ \nu$. Let $\mu_0>0$ be the vacuum permeability. The ideal MHD equations for $(\nu, u, B, h)$ on $\Omega$ are as follows: 
\sys{
&\nu_t+\mf{L}_{\bol{u}}\nu=0,\label{nut}\\
&u_t+\mf{L}_{\bol{u}}u+dh=
-\frac{1}{\rho}\iota_{\bol{J}}B 
,\label{ut}\\
&B_t+\mf{L}_{\bol{u}}B=0,\label{Bt}\\
&
\mu_0\bol{J}=\lr{\delta B}^{\sharp}
,\label{J0}\\
&dB=0.\label{dB0}
}{iMHD}
If $\rho\neq \textrm{const.}$, $h+\frac{1}{2}|\bol{u}|^{2}$ is a function of $\rho$.
\end{mydef}

\begin{remark}[The Lie--advection equation]
The equations \eqref{nut} and \eqref{Bt} represent the Lie--advection equations for the $n$-form $\nu$ and the $2$-form $B$, respectively. 
Differentiating \eqref{ut} yields the vorticity equation for the $2$-form $w = du$:
\begin{align}
w_t + \mf{L}_{\bol{u}} w = -d\!\left(\frac{1}{\rho}\,\iota_{\bol{J}} B\right). \label{wt}
\end{align}
If the Lorentz force vanishes, the vorticity also evolves according to a Lie--advection equation. 
In three-dimensional Euclidean space, (i) a vortex moves with the fluid particles (the frozen-in law), and (ii) the flux of vorticity across any surface moving with the fluid remains constant in time (Kelvin's circulation theorem); see, e.g., \cite[Theorems~1.21--1.22]{BV}. 
These two properties (Helmholtz's first and second laws, e.g., \cite[§3.4.1]{Davidson}) follow from the Lie--advection equation for the $2$-form $w$~\cite{ArnoldKh}. 
Analogous properties for the magnetic field are known as Alfvén's theorem~\cite[§4.3.1]{Davidson}. 
The Helmholtz laws (i) and (ii) can be generalized to differential $2$-forms $w$ and $B$ satisfying the Lie--advection equations on higher-dimensional manifolds. 
For instance, the second law (ii) for $w$ also holds on a higher-dimensional manifold in the sense that
\begin{equation*}
\frac{d}{dt} \int_{S(t)} w 
= \int_{S(t)} \left( \frac{\partial w}{\partial t} + \mf{L}_{\bol{u}} w \right) 
= 0,
\end{equation*}
where \( S(t) \) is a two-dimensional region comoving with the fluid in \( \Omega \); see, e.g., \cite[p.~142, Theorem~4.42]{TF}.
\end{remark}


\begin{remark}[The momentum balance]
The equation \eqref{ut} is a momentum balance on a manifold. By using 
\begin{equation*}{\mf{L}_{\bol{u}}u=\nabla_{\bol{u}} u+\frac{1}{2}d\abs{\bol{u}}^2=\lr{\nabla_{\bol{u}}\bol{u}}^{\flat}+\frac{1}{2}d\abs{\bol{u}}^2},
\end{equation*}
with $\abs{\bol{u}}^2=u_iu^i$, the $1$-form $\lr{\nabla_{\bol{u}}\bol{u}}^{\flat}$ corresponding to  the vector field $\nabla_{\bol{u}}\bol{u}$, and taking $\sharp$ of equation \eqref{ut}, the momentum balance can be expressed by vector fields as
\begin{equation*}
\bol{u}_t+\nabla_{\bol{u}}\bol{u}+\nabla\lr{h+\frac{1}{2}\abs{\bol{u}}^2}=-\frac{1}{\rho}J^iB_{ij}g^{jk}\p_k.\label{ut2}
\end{equation*} 
The usual form of the equation \eqref{ut} in Euclidean space follows from $\nabla_{\bol{u}}\bol{u}=\bol{u}\cdot \nabla \bol{u}$.
\end{remark}

\begin{remark}[The Amp\`ere's law]
The equation \eqref{J0} is the Amp\`ere's law in a curved space. Suppose that for an arbitrary $n$-dimensional volume $V_0\subset\Omega$,  
\begin{equation*}
\mu_0\int_{V_0}J^i\,dV=\int_{\p V_0}B^{ij}n_j\,dS,
\end{equation*} 
where $dV=\sqrt{g}\,dx^1\w ...\w dx^n$ and $dS=
\iota_{\bol{n}}dV$ denote volume and surface elements, and $n_j$ denotes the components of the unit outward normal $\bol{n}$ to the boundary $\p V_0$. By integration by parts, the component expression of \eqref{J0} follows: 
\begin{align*}
\mu_0J^i=\nabla_jB^{ij}.
\end{align*}
The usual relationship between $\bol{J}$ and $B$ in Euclidean space $\mu_0J^i= \partial_{j}B_{ij}$ follows from $\nabla_j B^{ij}=g^{ij}\nabla^k B_{jk}$. 
\end{remark}


\begin{remark}[The closure condition]
The equation \eqref{dB0} is an intrinsic generalization of the three-dimensional divergence-free condition on the magnetic field. An alternative higher-dimensional generalization is to extend the magnetic field $\bol{B}$ as an $n$-dimensional divergence-free vector field. For such magnetic fields, the induction equation \eqref{iMHDBt} can be generalized as 
\begin{equation*}
\frac{\p\bol{B}}{\p t}+\lrs{\bol{u},\bol{B}}=0,
\end{equation*}
where $\lrs{\bol{u},\bol{B}}=\bol{u}\cdot\nabla\bol{B}-\bol{B}\cdot\nabla\bol{u}$ denotes the Lie-bracket of vector fields. The current field \eqref{iMHDBt} and the momentum equations \eqref{iMHDut} can be adjusted as $\bol{J}=(\nabla \bol{B}-{}^{t}(\nabla \bol{B}))/2$ with the Lorentz force of the form $\frac{1}{\mu_0}(\bol{B}\cdot\nabla\bol{B}-\frac{1}{2}\nabla\abs{\bol{B}}^2)$. The other equations in \eqref{iMHD3} remain unchanged. It is unknown whether such an extrinsic generalization admits higher-dimensional topological invariants such as \eqref{eq: C}, \eqref{eq: M}, and \eqref{eq: W}.
\end{remark}

\begin{remark}[Relativistic MHD]
The system~\eqref{iMHD} provides a geometric generalization of ideal MHD, extending the formulation for three-dimensional vector fields in Euclidean space to one for differential forms on higher-dimensional Riemannian manifolds. It reproduces ideal hydrodynamics on a Riemannian manifold in the case of constant-density, e.g., \cite{ArnoldKh}, and classical ideal MHD in three-dimensional Euclidean space~\cite[pp.~107--108]{Fitz}, \cite[p.~7]{Freid}. 
The differential forms encoding physical quantities and the underlying geometric (Lie-advection) structure are consistent with those appearing in ideal relativistic hydrodynamics~\cite[pp.~82--83]{Frankel79}, \cite[pp.~47--52 and 588--597]{Wein72}, and in relativistic MHD~\cite{DAvignon15,Marsden86,Kaw17,Bau03} in four space-time dimensions. 
\end{remark}

\subsection{The Hodge--Morrey decomposition}

We describe the magnetic fields and their vector potentials in the intrinsic system~\eqref{iMHD} by applying the Hodge--Morrey decomposition. 
Let $\Lambda^{k}(\Omega)$ denote the space of all smooth $k$-forms on $\Omega$ for $0\leq k\leq n$. The $k$-form $\omega$ is normal on $\partial\Omega$ (resp. tangential) if $i^{*}\omega=0$ on $\partial\Omega$ (resp. $i^{*}*\omega=0$ on $\partial\Omega$) for the inclusion map $i: \partial\Omega\hookrightarrow \Omega$ and its pullback $i^{*}:\Lambda^{k}(\Omega)\to \Lambda^{k}(\partial\Omega)$. We denote by $\Lambda_{D}^{k}(\Omega)$ (resp. $\Lambda_{N}^{k}(\Omega)$) the space of all elements in $\Lambda^{k}(\Omega)$ normal on $\partial\Omega$ (resp. tangential on $\partial\Omega$). We denote the space of all harmonic $k$-forms on $\Omega$ by $\mathcal{H}^{k}(\Omega)=\{\gamma\in \Lambda^k(\Omega)\ |\ d\gamma=0,\ \delta \gamma=0\}$. 

We denote by $L^{2}\Lambda^{k}(\Omega)$ the completion of $\Lambda^{k}(\Omega)$ with the $L^{2}$-norm induced by the inner product
\begin{align*}
<\omega,\eta>=\int_{\Omega}\omega\wedge *\eta.
\end{align*}
Similarly, we define $H^{1}\Lambda^{k}_{D}(\Omega)$ and $H^{1}\Lambda^{k}_{N}(\Omega)$ by the completions of $\Lambda^{k}_D(\Omega)$ and $\Lambda^{k}_N(\Omega)$ with the $H^{1}$-norm, respectively. We apply the following Hodge--Morrey decomposition \cite[Theorem 2.4.2]{Schwarz}.

\begin{lemma}[Hodge--Morrey decomposition]\label{l: HM}
Let $\Omega$ be a compact and orientable $n$-dimensional Riemannian manifold with smooth boundary for $n\geq 2$. Then, for $0<k<n$, 
\begin{align}
L^{2}\Lambda^{k}(\Omega)=d H^{1}\Lambda_D^{k-1}(\Omega)\oplus \delta H^{1}\Lambda_N^{k+1}(\Omega)\oplus \mathcal{H}^{k}(\Omega).  \label{eq: HM}
\end{align}
The space $\mathcal{H}^{k}(\Omega)$ is isomorphic to the deRham cohomology group $\bol{H}^{k}(\Omega,d)=\textrm{Ker}\ d|_{\Lambda^{k}(\Omega)}/\textrm{Im}\ d|_{\Lambda^{k-1}(\Omega)}$. In particular, $\textrm{dim}\ \mathcal{H}^{k}(\Omega)$ agrees with the Betti number $b_k(\Omega)=\textrm{dim}\ \bol{H}^{k}(\Omega,d)$.
\end{lemma}

\begin{example}
When $\Omega=\mathbb{S}^{n}$ and $\mathbb{T}^{n}$, 
\begin{align*}
b_k(\mathbb{S}^{n})=\begin{cases}
\ 1 & k=0,n,\\
\ 0 & 0<k<n,
\end{cases} \quad 
b_k(\mathbb{T}^{n})=\binom{n}{k}.
\end{align*}
When $\Omega\subset \mathbb{R}^{2}$ is a bounded multiply-connected domain with genus $g$, 
\begin{align*}
b_k(\Omega)=\begin{cases}
\ 1 & k=0,\\
\ g & k=1,\\
\ 0 & k=2.
\end{cases}
\end{align*}
When $\Omega\subset \mathbb{R}^{3}$ is a bounded multiply-connected domain with genus $g$ and cavity $h$, 
\begin{align*}
b_k(\Omega)=\begin{cases}
\ 1 & k=0,\\
\ g & k=1,\\
\ h & k=2,\\
\ 0 & k=3.
\end{cases}
\end{align*}
\end{example}

\subsection{MHS equilibria}

Steady solutions of the ideal MHD equations~\eqref{iMHD} describe higher-dimensional MHS equilibria.

\begin{mydef}[MHS equilibria] 
We say that $(B,h)$ is an MHS equilibrium in $\Omega$ for constant $\rho_0$ if $(B,h)$ satisfy  the following: 
\sys{
& dh=
-\frac{1}{\rho_0}\iota_{\bol{J}}B,\label{ut2}\\
&
\mu_0\bol{J}=
\lr{\delta B}^{\sharp},\label{J02}\\
&dB=0.\label{dB0b}
}{MHS}
\end{mydef}

\begin{remark}[Potential form MHS]
The stationary Euler equations and the intrinsic MHS equilibrium equations~\eqref{MHS} coincide in three dimensions; see Table~\ref{t: EMHS}. 
In higher dimensions, the stationary Euler equations comprise \( n + 1 \) equations for \( n + 1 \) unknowns \((u, \Pi)\). 
In contrast, the intrinsic MHS equilibrium equations~\eqref{MHS} consist of \( n + \binom{n}{3} \) equations, given by the force balance~\eqref{ut2} and the closure condition~\eqref{dB0b}, with \( \binom{n}{2} + 1 \) unknowns \((B, h)\), representing the components of the \(2\)-form \(B\) and the scalar pressure \(h\). 
These two numbers coincide if and only if \( n = 3 \).


This discrepancy can be reconciled by introducing a $1$-form expressing the MHS equilibrium \eqref{MHS} for $B$. Indeed, for an MHS equilibrium $(B,h)$, there exist a $1$-form $A\in 
H^{1}\Lambda_D^{1}(\Omega)$ and 
a harmonic $2$-form 
$\xi\in \mathcal{H}^{2}(\Omega)$ such that $B=dA+\xi$ by the Hodge--Morrey decomposition. We choose the Coulomb gauge $\delta A=0$ so that $A$ is unique up to elements of $\mathcal{H}^{1}(\Omega)$. The $1$-form $A$ then satisfies the equations 
\begin{align*}
& dh = -\frac{1}{\rho_0}\,\iota_{\bol{J}}(dA+\xi), \\ 
& \mu_0 \bol{J} = \lr{\Delta A}^{\sharp},  \\
& \delta A=0,
\end{align*}
for $\Delta=d\delta +\delta d$. Conversely, for a solution $(A,h)$ to this system for a given harmonic $2$-form $\xi \in \mathcal{H}^{2}(\Omega)$, $B=dA+\xi$ is an MHS equilibrium \eqref{MHS}. The system for the $1$-form comprises $n+1$ equations for the $n+1$ unknowns $(A,h)$.
\end{remark}

\begin{remark}[Higher-dimensional MHS equilibria]
The topology of higher-dimensional MHS equilibria \eqref{MHS} and their existence/non-existence are largely unexplored. We recall known results for higher-dimensional steady Euler flows \cite{ArnoldKh}:
\begin{itemize}
\item (Generalized Beltrami flows)
On an odd-dimensional manifold $\Omega$ with dimension $2m+1$, a trajectory of an analytic divergence-free vector field $\bol{u}$ is called chaotic if it is not contained on any analytic hypersurface in $\Omega$. A vorticity field $\bol{\omega}$ is a vector field such that $\iota_{\bol{\omega}}dV=w^{m}$ for the vorticity form $w=du$ and $u=\bol{u}^{\flat}$. The vector field $\bol{\omega}$ is called a generalized Belrami flow if there exists a constant $C$ such that $\bol{\omega}=C\bol{u}$ \cite[II. 6.A]{ArnoldKh}. An example of a generalized Beltrami flow without chaotic trajectories is the Hopf vector field $\bol{\omega}=(x^2,-x^1,x^4,-x^3,\cdots,x^{2m+2},-x^{2m+1})\in \mathbb{S}^{2m+1}$. Ghrist \cite{Gh01} constructed a generalized Beltrami flow with chaotic trajectories on a manifold by using a metric. Those generalized Beltrami flows appear to be the only known examples of higher-dimensional steady Euler flows.

\item (Four-dimensional steady Euler flows)
On an even-dimensional manifold $\Omega$ with dimension $2m$, the vorticity function $\lambda=w^{m}/dV$ is an additional first integral of the velocity field $\bol{u}$ of the steady Euler flows besides the Bernoulli function $\Pi$. Ginzburg and Khesin \cite{GK1}, \cite{GK2} showed that trajectories of the four-dimensional steady Euler flows are similar to an integrable Hamiltonian system with two degrees of freedom; see also \cite[II, 6.B]{ArnoldKh}. Similar to the vorticity function of the steady Euler flows, the function $\mu=B^{m}/dV$ of the MHS equilibrium $B$ is an additional first integral of the current field $\bol{J}$ besides the Bernoulli function $h$. 
It is a question whether four-dimensional MHS equilibria exhibit a similar topology of solutions. 
\end{itemize}
\end{remark}

\section{Even-dimensional invariants}

We show the conservation of the generalized enstrophy~\eqref{eq: En} for the vorticity of the inhomogeneous Euler equations, and of an analogous quantity~\eqref{eq: W} for the magnetic field of the ideal MHD system~\eqref{iMHD}, on a compact and orientable \( (2m) \)-dimensional Riemannian manifold \( \Omega \) with smooth boundary, for \( m \ge 1 \). 
These conservation laws follow from the Lie--advection equations for the vorticity and the magnetic field.  


\begin{prop}\label{p: CGE}
(i) The generalized fluid  enstrophy
\begin{align}
\mathscr{E}=\int_{\Omega}f\lr{\frac{w^{m}}{\nu}}\nu ,\label{en}
\end{align}
is constant for smooth solutions $(\nu,u,0,h)$ to \eqref{iMHD} and $w=du$ subject to the boundary condition $\bol{u}\cdot \bol{n}=0$ on $\partial\Omega$, for arbitrary functions $f$.\\
\noindent
(ii) The quantity 
\begin{align}
\mathscr{W}=\int_{\Omega}f\lr{\frac{B^{m}}{\nu}}\nu ,\label{aen}
\end{align}
is constant for smooth solutions $(\nu,u,B,h)$ to \eqref{iMHD} subject to the boundary condition $\bol{u}\cdot \bol{n}=0$ on $\partial\Omega$, for arbitrary functions $f$.
\end{prop}

\begin{proof}
(i) We express the ($2m$)-form $w^{m}$ as $w^m=W\,dV$. By the vorticity equation \eqref{wt}, $w_t^{m}+\mf{L}_{\bol{u}}w^{m}=0$. By using $d\iota_{\bol{u}}\lr{W\,dV}=(\textrm{div}\ \lr{W\bol{u}})\,dV$, we obtain  
\begin{align*}
W_t+\textrm{div}(W\bol{u})=0.
\end{align*}
Similarly, the ($2m$)-form $\nu =\rho\, dV$ satisfies $\rho_t+\textrm{div}(\rho\bol{u})=0$. It follows that 
\begin{align*}
\left(\frac{W}{\rho}\right)_t+
\mf{L}_{\bol{u}}\left(\frac{W}{\rho}\right)=0. 
\end{align*} 
Thus, for an arbitrary $f\lr{w^m/\nu}=f\lr{W/\rho}$,
\begin{align*}
f
_t+
\mf{L}_{\bol{u}}f
=0. 
\end{align*} 
By using the continuity equation $\nu_t=-\mf{L}_{\bol{u}}\nu$, 
\begin{align*}
\lr{f\nu}_t+d\iota_{\bol{u}}\left(  f \nu \right)=0. 
\end{align*}
Thus, the conservation of \eqref{en} follows from Stokes's theorem. 

(ii) The same computation using the Lie-advection equation \eqref{iMHDBt} yields the conservation of \eqref{aen}.
\end{proof}

\section{Odd-dimensional invariants}

We show the conservation of the fluid helicity~\eqref{eq: He}, cross helicity~\eqref{eq: C}, and magnetic helicity~\eqref{eq: M} for smooth solutions of the ideal MHD system~\eqref{iMHD}, thereby completing the proof of Theorem~\ref{t: thm}. 
In contrast to the even-dimensional invariants~\eqref{en} and~\eqref{aen}, which follow from the Lie--advection equations for $2$-forms, the odd-dimensional invariants are governed by the $1$-form equations for the fluid velocity and the magnetic potential; see Table~\ref{t: T3}. 
We derive $(2m+1)$-form equations by combining these two equations and establish conservation by applying Stokes's theorem under the boundary conditions $\bol{u}\!\cdot\!\bol{n}=0$ and $i^{*}w=0$ or $i^{*}B=0$ on~$\partial\Omega$.


\begin{table}[h]
\centering
\begin{tabular}{cccc}
\hline
\multicolumn{1}{|c|}{}                                                                        & \multicolumn{1}{c|}{\begin{tabular}[c]{@{}c@{}}$2$-form equations\end{tabular}} & \multicolumn{1}{c|}{$1$-form equations}                              & \multicolumn{1}{c|}{($2m+1$)-form equations}                                                              \\ \hline
\multicolumn{1}{|c|}{\begin{tabular}[c]{@{}c@{}}Generalized \\ Fluid helicity\end{tabular}}   & \multicolumn{1}{c|}{$w_t+\mf{L}_{\bol{u}}w=0$}                                          & \multicolumn{1}{c|}{$u_t+
\mf{L}_{\bol{u}}u+dh
=0$} & \multicolumn{1}{c|}{$(u\wedge w^{m})_t+d (\iota_{\bol{u}}(u\wedge w^{m})+hw^{m}  )=0$}       \\ \hline
\multicolumn{1}{|c|}{\begin{tabular}[c]{@{}c@{}}Generalized \\ Cross helicity\end{tabular}}   & \multicolumn{1}{c|}{$B_t+\mf{L}_{\bol{u}}B=0$}                                          & \multicolumn{1}{c|}{$u_t+
\mf{L}_{\bol{u}}u+dh
=-\frac{1}{\rho}\iota_{\bol{J}}B$} & \multicolumn{1}{c|}{$(u\wedge B^{m})_t+d (\iota_{\bol{u}}(u\wedge B^{m})+hB^{m}  )=0$}       \\ \hline
\multicolumn{1}{|c|}{\begin{tabular}[c]{@{}c@{}}Generalized\\ Magnetic helicity\end{tabular}} & \multicolumn{1}{c|}{$B_t+\mf{L}_{\bol{u}}B=0$}                                          & \multicolumn{1}{c|}{$A_t+\iota_{\bol{u}}B+d\Phi=0$}                  & \multicolumn{1}{c|}{$(A\wedge B^{m})_{t}+d(-A\wedge\iota_{\bol{u}}B^{m}+\Phi  B^{m})=0$} \\ \hline
\multicolumn{1}{l}{}                                                                          & \multicolumn{1}{l}{}                                                                    & \multicolumn{1}{l}{}                                                 & \multicolumn{1}{l}{}                                                                        
\end{tabular}
\caption{Odd-dimensional invariants and the originating differential forms}\label{t: T3}
\end{table}

Throughout this section, we assume that $\Omega$ is a compact and orientable $(2m+1)$-dimensional Riemannian manifold with smooth boundary $\partial\Omega$ for $m\geq 1$. 

\subsection{Fluid helicity and cross helicity}

We show conservation of fluid helicity \eqref{eq: He} and cross helicity \eqref{eq: C} by deriving ($2m+1$)-form equations for $u\wedge w^{m}$ and $u\wedge B^{m}$.

\begin{prop}\label{p: CHC}
(i) The fluid helicity 
\eq{
\mathscr{H}=\int_{\Omega}u\w w^{m},\label{H} 
}
is constant for smooth solutions $(\nu,u,0,h)$ to \eqref{iMHD} for the vorticity form $w=du$ satisfying the boundary conditions  $\bol{u}\cdot \bol{n}=0$ and $i^{*}w=0$ on $\partial\Omega$.\\
\noindent
(ii) The cross helicity
\eq{
\mathscr{C}=\int_{\Omega} u\wedge B^m,\label{C}
}
is constant for smooth solutions $(\nu,u,B,h)$ to \eqref{iMHD} subject to the boundary conditions  $\bol{u}\cdot \bol{n}=0$ and $i^{*}B=0$ on $\partial\Omega$.
\end{prop}

\begin{proof}
(i) We show the identity 
\begin{align}
(u\wedge w^{m})_t+d (\iota_{\bol{u}}(u\wedge w^{m})+hw^{m}  )=0,  \label{eq: DH}
\end{align}
for smooth solutions $(\nu,u,0,h)$ to \eqref{iMHD}. The conservation of \eqref{H} then follows from Stokes's theorem. By the momentum balance \eqref{ut}, $u_t+\iota_{\bol{u}}w+d(\iota_{\bol{u}}u+h ) =0$. We take the exterior product with $w^{m}$. Since the ($2m+2$)-form $w^{m+1}$ vanishes, $\iota_{\bol{u}}w\wedge w^{m}=\frac{1}{m+1}\iota_{\bol{u}}w^{m+1}=0$ and 
\begin{align*}
u_t\wedge w^{m}+d\iota_{\bol{u}}u \wedge w^{m}+d(hw^{m})=0.
\end{align*}
By the vorticity equation  \eqref{wt}, $w_t^{m}+d\iota_{\bol{u}}w^{m}=0$. By taking the exterior product with $u$, 
\begin{align*}
u\wedge w^{m}_t+u\wedge d\iota_{\bol{u}}w^{m}=0.
\end{align*}
By computation,  
\begin{align*}
d (\iota_{\bol{u}}(u\wedge w^{m})  )
=d(\iota_{\bol{u}}u\wedge w^{m}-u\wedge \iota_{\bol{u}} w^{m}  )
&=d\iota_{\bol{u}}u\wedge w^{m}-w\wedge \iota_{\bol{u}} w^{m}+u\wedge d \iota_{\bol{u}} w^{m} \\
&=d\iota_{\bol{u}}u\wedge w^{m}+u\wedge d \iota_{\bol{u}} w^{m},
\end{align*}
and \eqref{eq: DH} follows.

(ii) We show the identity 
\begin{align}
 (u\wedge B^{m})_t+d (\iota_{\bol{u}}(u\wedge B^{m})+hB^{m}  )=0,  \label{eq: DC}
\end{align}
for smooth solutions $(\nu,u,B,h)$ to \eqref{iMHD}. By the momentum balance \eqref{ut}, $u_t+\iota_{\bol{u}}du+d\iota_{\bol{u}}u+dh =-\frac{1}{\rho}\iota_{\bol{J}}B$. We take the exterior product with the  ($2m$)-form $B^{m}$. Since the ($2m+2$)-forms $d u\wedge B^{m}$ and $B^{m+1}$ vanish and $dB=0$,
\begin{align*}
\iota_{\bol{u}}du\wedge B^{m}&=\iota_{\bol{u}}(du\wedge B^{m} )-du\wedge \iota_{\bol{u}}B^{m}=-du\wedge \iota_{\bol{u}}B^{m},\\
\frac{1}{\rho}\iota_{\bol{J}}B\wedge B^{m}&=\frac{1}{(m+1)\rho}\iota_{\bol{J}}B^{m+1}=0.
\end{align*}
We thus obtain 
\begin{align*}
u_t\wedge B^{m}-du\wedge \iota_{\bol{u}}B^{m}+d\iota_{\bol{u}}u\wedge B^{m}+d(h B^{m})  =0.
\end{align*}
By the induction equation  \eqref{Bt}, $B^{m}_{t}+d\iota_{\bol{u}}B^{m}=0$. By taking the exterior product with $u$,
\begin{align*}
u\wedge B^{m}_t+u\wedge d \iota_{\bol{u}}B^{m}=0.
\end{align*}
By computation,  
\begin{align*}
d\iota_{\bol{u}}(u\wedge B^{m})
=d(\iota_{\bol{u}}u\wedge B^{m}-u\wedge \iota_{\bol{u}} B^{m} )
=d\iota_{\bol{u}}u\wedge B^{m}-du\wedge \iota_{\bol{u}} B^{m}
+u\wedge d\iota_{\bol{u}} B^{m},
\end{align*}
and \eqref{eq: DC} follows.
\end{proof}

\subsection{Magnetic helicity}

We show the conservation of magnetic helicity~\eqref{eq: M} when the cohomology groups of \( \Omega \) are trivial, thereby completing the proof of Theorem~\ref{t: thm}.

\begin{prop}\label{p: HM1}
Assume that $b_k=0$ for $k=1$ and $2$. Then, for a closed $2$-form $B\in L^{2}\Lambda^{2}(\Omega)$, there exists a $1$-form $A\in H^{1}\Lambda^{1}_{D}(\Omega)$ such that $B=dA$. The $1$-form $A$ is unique up to elements of $dH^{1}\Lambda^{0}_{D}(\Omega)$.
\end{prop}

\begin{proof}
For a closed $2$-form $B\in L^{2}\Lambda^{2}(\Omega)$, there exists a $1$-form $A\in H^{1}\Lambda^{1}_{D}(\Omega)$ and $3$-form $C\in H^{3}\Lambda^{1}_{N}(\Omega)$ such that $B=dA+\delta C$ by Hodge--Morrey decomposition (Lemma \ref{l: HM}). Since $dB=0$, $\delta C$ is a harmonic 2-form and hence $\delta C=0$. For $\tilde{A}\in H^{1}\Lambda^{1}_{D}(\Omega)$ such that $B=d\tilde{A}$, there exist $\alpha\in H^{1}\Lambda^{0}_{D}(\Omega)$ and $\beta\in H^{1}\Lambda^{2}_{N}(\Omega)$ such that $\tilde{A}-A=d\alpha +\delta \beta$ by Hodge--Morrey decomposition. By $d(\tilde{A}-A)=0$, $\delta \beta$ is a harmonic $1$-form and hence $\delta \beta=0$. Thus $A$ is unique up to 
exact $1$-forms.
\end{proof}

\begin{prop}
Let $B\in L^{2}\Lambda^{2}(\Omega)$ be a smooth closed $2$-form. Let $A\in H^{1}\Lambda_D^{1}(\Omega)$ be a $1$-form such that $B=dA$. Then, the value of the integral  
\begin{equation*}
\int_{\Omega} A\wedge B^m,
\end{equation*}
is independent of the choice of $A$.
\end{prop}

\begin{proof}
Let $A$, $\tilde{A}\in H^{1}\Lambda_D^{1}(\Omega)$ be two $1$-forms such that $B=dA=d\tilde{A}$. Then, $\tilde{A}=A+d\alpha$ for some $\alpha\in H^{1}\Lambda_D(\Omega)$ by Proposition \ref{p: HM1}. Then, 
\begin{align*}
\int_{\Omega} \tilde{A}\wedge B^m=\int_{\Omega} A\wedge B^m+\int_{\Omega} d\alpha \wedge B^m.
\end{align*}
By $d(\alpha\wedge B^{m})=d\alpha\wedge B^{m}$ and $i^{*}\alpha=0$ on $\partial\Omega$, the last integral vanishes by Stokes's theorem.
\end{proof}

\begin{mydef}(Magnetic helicity with $b_1=b_2=0$) 
Let $\Omega$ be a compact, orientable $(2m+1)$-dimensional Riemannian manifold with smooth boundary $\partial\Omega$ for $m\geq 1$ with $b_k=0$ for $k=1$ and $2$. For a smooth closed 2-form $B\in L^{2}\Lambda^{2}(\Omega)$, we define magnetic helicity 
\eq{
\mathscr{M} = \int_{\Omega} A \wedge B^{m},\label{MH3}
}
where $A\in H^{1}\Lambda^{1}_{D}(\Omega)$ is a $1$-form such that $B=dA$.  
\end{mydef}

\begin{prop}\label{p: MHC}
The magnetic helicity~\eqref{MH3} is constant for smooth solutions \((\nu, u, B, h)\) to~\eqref{iMHD}
subject to the boundary conditions 
\(\bol{u}\!\cdot\!\bol{n} = 0\) and \( i^{*}B = 0 \) on~\(\partial\Omega\). 
\end{prop}

\begin{proof}
We take a $1$-form $A\in H^{1}\Lambda_D^{1}(\Omega)$ such that $B=dA$ by Proposition \ref{p: HM1}. By the induction equation  \eqref{Bt}, $d(A_t+\iota_{\bol{u}}B)=0$. By Hodge--Morrey decomposition, there exists $\Phi\in H^{1}\Lambda_D^{0}(\Omega)$ such that $A_t+\iota_{\bol{u}}B+d\Phi=0$. By taking the exterior product with $B^{m}$, 
\begin{align*}
A_t\wedge B^{m}+d(\Phi  B^{m})=0.
\end{align*}
By $B^{m}_t+d\iota_{\bol{u}}B^{m}=0$ and $d(A\wedge \iota_{\bol{u}}B^{m})=-A\wedge d\iota_{\bol{u}}B^{m}$, 
\begin{align*}
A\wedge B^{m}_{t}-d(A\wedge \iota_{\bol{u}}B^{m})=0.
\end{align*}
We obtain 
\begin{align*}
(A\wedge B^{m})_{t}+d(-A\wedge\iota_{\bol{u}}B^{m}+\Phi\wedge B^{m})=0,
\end{align*}
and the conservation of \eqref{MH2} follows from Stokes's theorem.
\end{proof}

\begin{proof}[Proof of Theorem \ref{t: thm}]
The results follow from Propositions \ref{p: CGE}, \ref{p: CHC}, and \ref{p: MHC}.
\end{proof}

\subsection{Extensions to manifolds with non-zero Betti numbers}


We extend the magnetic helicity conservation result of Theorem~\ref{t: thm} to manifolds \( \Omega \) with nonzero Betti numbers \( b_k \neq 0 \) for \( k = 1, 2 \), and to solutions of~\eqref{iMHD} with exact initial magnetic field \(2\)-forms.

\subsubsection{The orthogonal condition}

When $\Omega$ is a bounded and multiply-connected domain in $\mathbb{R}^{3}$ (with $b_{1}\neq 0$ and $b_{2}=0$), magnetic helicity \eqref{eq: MH} is gauge-invariant for the divergence-free vector field $\bol{B}$ satisfying $\bol{B}\cdot \bol{n}=0$ on $\partial \Omega$ and the orthgonal condition
\begin{align}
\int_{\Omega}\bol{\gamma}_j\cdot \bol{B}\,dV=0,\quad j\in \{1,\cdots,b_{1}\},  \label{eq: FC}
\end{align}
where $\{\bol{\gamma}_j\}_{j=1}^{b_{1}}$ is a basis of harmonic vector fields on $\Omega$. Indeed, for any $\bol{A}$ and $\bol{\tilde{A}}$ satisfying $\nabla \times \bol{A}=\bol{B}$ and $\nabla \times \bol{\tilde{A}}=\bol{B}$, $\bol{\tilde{A}}=\bol{A}+\nabla \alpha+\bol{\gamma}$ with some scalar function $\alpha$ and a harmonic vector field $\bol{\gamma}$. Thus, $\mathscr{M}=\int_{\Omega}\bol{A}\cdot\bol{B}\,dV=\int_{\Omega}\bol{\tilde{A}}\cdot\bol{B}\,dV$ by the orthogonal condition \eqref{eq: FC}. The condition \eqref{eq: FC} is equivalent to the flux condition 
\begin{align}
\int_{\Sigma_j}\bol{B}\cdot\bol{n}_j\,d S=0,\quad j\in \{1,\cdots,b_{1}\},  \label{FC2}
\end{align}
for cutting surfaces $\{\Sigma_{j}\}_{j=1}^{b_{1}}$ of $\Omega$; see \cite[p. 6]{TV}. MacTaggart and Valli \cite[eq. (2.20)]{TV} derived a gauge-invariant magnetic helicity
\eq{
\mathscr{M}
 = \int_{\Omega} \bol{A}\cdot\bol{B}\,dV
   - \sum_{j=1}^g 
     \left( \int_{C_j} \bol{A}\cdot\bol{t}_jdl \right)
     \left( \int_{\Sigma_j} \bol{B}\cdot\bol{n}_j dS\right),
\label{TVH}
} 
where $C_j=\partial \Sigma_j'$ are closed curves for cutting surfaces  $\{\Sigma_j'\}_{j=1}^{g}$ of $\Omega'=B\backslash \overline{\Omega}$ with an open ball $B$ containing $\Omega$. The formula \eqref{TVH} generalizes the Bevir--Gray formula \cite{BG} for $b_1=1$ and applies to all divergence-free $\bol{B}$ satisfying $\bol{B}\cdot \bol{n}=0$ on $\partial\Omega$ with a vector potential $\bol{A}$ such that $\nabla \times \bol{A}=\bol{B}$. See also \cite{Pfeff21}.

Let us define gauge-invariant magnetic helicity for higher-dimensional manifolds with $b_k\neq 0$ for $k=1,2$ by generalizing the orthogonal condition \eqref{eq: FC} for magnetic $2$-forms.  

\begin{prop} \label{p: HM3}
For a closed 2-form $B\in L^{2}\Lambda^{2}(\Omega)$, there exists a $1$-form $A\in H^{1}\Lambda^{1}_{D}(\Omega)$ and a harmonic 2-form $\xi\in \mathcal{H}^{2}(\Omega)$ such that
\begin{align}
B=dA+\xi.    \label{eq: BD}
\end{align}
The $1$-form $A$ is unique up to elements of $dH^{1}\Lambda^{0}_{D}(\Omega)\oplus \mathcal{H}^{1}(\Omega)$.
\end{prop}

\begin{proof}
By Hodge--Morrey decomposition (Lemma \ref{l: HM}), there exist a $1$-form $A\in H^{1}\Lambda^{1}_{D}(\Omega)$, a $3$-form $C\in H^{1}\Lambda^{3}_{N}(\Omega)$, and a harmonic 2-form $\xi\in \mathcal{H}^{2}(\Omega)$ such that
\begin{align*}
B=dA+\delta C+\xi.   
\end{align*}
By $dB=0$, $\delta C$ is a harmonic $2$-form, and hence $\delta C=0$ by uniqueness of the decomposition. Let $\tilde{A}\in H^{1}\Lambda^{1}_{D}(\Omega)$ be a $1$-form such that $B=d\tilde{A}+\xi$. Then, there exist a $0$-form $\alpha\in H^{1}\Lambda^{0}_{D}(\Omega)$, a $2$-form $\beta\in H^{1}\Lambda^{3}_{N}(\Omega)$, and a harmonic $1$-form $\gamma\in \mathcal{H}^{1}(\Omega)$ such that $\tilde{A}-A=d\alpha+\delta \beta+\gamma$. By $d (\tilde{A}-A)=0$, $\delta\beta$ is a harmonic $1$-form and hence $\delta\beta=0$ by uniqueness of the decomposition. Thus $\tilde{A}=A+d \alpha+\gamma$ and $A$ is unique up to elements of $dH^{1}\Lambda^{0}_{D}(\Omega)\oplus \mathcal{H}^{1}(\Omega)$. 
\end{proof}

\begin{prop}
Let $B\in L^{2}\Lambda^{2}(\Omega)$ be a smooth closed $2$-form satisfying the orthogonal condition 
\begin{align}
\int_{\Omega}\gamma_j\wedge B^{m}=0\quad j\in \{1,\cdots,b_1\},    \label{eq: FC3}
\end{align}
for a basis $\{\gamma_j\}_{j=1}^{b_1}$ of $\mathcal{H}^{1}(\Omega)$. Let $A\in H^{1}\Lambda_D^{1}(\Omega)$ be a $1$-form satisfying \eqref{eq: BD}. Then, the constant 
\begin{align*}
\int_{\Omega} A\wedge B^m,\label{CH}
\end{align*}
is independent of the choice of $A$.
\end{prop}

\begin{proof}
Let $A$, $\tilde{A}\in H^{1}\Lambda_D^{1}(\Omega)$ be two $1$-forms satisfying \eqref{eq: BD}. Then, $\tilde{A}=A+d\alpha+\gamma$ for some $\alpha\in H^{1}\Lambda_D(\Omega)$ and $\gamma\in \mathcal{H}^{1}(\Omega)$ by Proposition \ref{p: HM3}. Then, 
\begin{align*}
\int_{\Omega} \tilde{A}\wedge B^m=\int_{\Omega} A\wedge B^m+\int_{\Omega} d\alpha \wedge B^m+\int_{\Omega} \gamma \wedge B^m.
\end{align*}
The second term on the right-hand side vanishes by Stokes's theorem. The last term also vanishes by the orthogonal condition \eqref{eq: FC3}.    
\end{proof}

\begin{mydef}(Magnetic helicity wth $b_1,b_2\neq 0$) 
Let $\Omega$ be a compact and orientable $(2m+1)$-dimensional Riemannian manifold with smooth boundary $\partial\Omega$ for $m\geq 1$. Let $\{\gamma_j\}_{j=1}^{b_1}$ be a basis of $\mathcal{H}^{1}(\Omega)$. For a smooth closed $2$-form $B\in L^{2}\Lambda^{2}(\Omega)$ satisfying the orthogonal condition \eqref{eq: FC3}, we define magnetic helicity 
\eq{
\mathscr{M} = \int_{\Omega} A \wedge B^{m},\label{MH2}
}
by a $1$-form $A\in H^{1}\Lambda^{1}_D(\Omega)$ satisfying \eqref{eq: BD}.  
\end{mydef}

We show that the orthogonal condition \eqref{eq: FC3} is conserved in the evolution of the system \eqref{iMHD} subject to the boundary conditions $\bol{u}\cdot \bol{n}=0$ and 
$i^{\ast}B = 0$ on $\partial\Omega$.

\begin{prop}
Let $(\nu,u,B,h)$ be a smooth solution to \eqref{iMHD} satisfying the boundary conditions $\bol{u}\cdot \bol{n}=0$ and $i^{\ast}B = 0$ on $\partial\Omega$. Assume that the initial magnetic field $2$-form $B_0$ satisfies the orthogonal condition \eqref{eq: FC3}. Then, \eqref{eq: FC3} holds for all time.  
\end{prop}

\begin{proof}
By the induction equations \eqref{Bt}, $B_t^{m}+d\iota_{\bol{u}}B^{m}=0$ and 
\begin{align*}
    \gamma_{j}\wedge B_t^{m}-d(\gamma_{j}\wedge \iota_{\bol{u}} B^{m})=0.
\end{align*}
By the boundary conditions $\bol{u}\cdot \bol{
n}=0$ and $i^{*}B=0$ on $\partial\Omega$ and Stokes's theorem, the property \eqref{eq: FC3} follows.
\end{proof}

\subsubsection{The exact form}

We show conservation of magnetic helicity \eqref{MH2} for solutions to \eqref{iMHD} and exact initial magnetic fields.

\begin{prop}\label{p: ex}
Exact $2$-forms $B=dA\in H^{1}\Lambda_D^{1}(\Omega)$ satisfy the orthogonal condition \eqref{eq: FC3}. 
\end{prop}

\begin{proof}
By $\gamma_j\wedge B^{m}=-d(\gamma_j\wedge A\wedge B^{m-1})$ and Stokes's theorem,
\begin{align*}
\int_{\Omega}\gamma_j\wedge B^{m}=-\int_{\Omega}d(\gamma_j\wedge A\wedge B^{m-1})=-\int_{\partial\Omega}i^{*}(\gamma_j\wedge A\wedge B^{m-1}).
\end{align*}
The last term vanishes by the boundary condition $i^{*}A=0$ on $\partial\Omega$.
\end{proof}

\begin{prop}\label{p: ex2}
Let $(\nu,u,B,h)$ be a smooth solution to \eqref{iMHD} satisfying the boundary conditions $\bol{u}\cdot \bol{n}=0$ and $i^\ast B=0$ on $\partial\Omega$. Assume that $B_0\in dH^{1}\Lambda_D^{1}(\Omega)$. Then, $B\in dH^{1}\Lambda_D^{1}(\Omega)$ for all time.  
\end{prop}

\begin{proof}
Let $A\in H^{1}\Lambda^{1}_{D}(\Omega)$ and $\xi\in \mathcal{H}^{2}(\Omega)$ satisfy \eqref{eq: BD} for a smooth solution $B\in L^{2}\Lambda^{2}(\Omega)$ to \eqref{iMHD}. By the boundary conditions $\bol{u}\cdot \bol{n}=0$ and $i^{*}B=0$ on $\partial\Omega$, $i^{*}\iota_{\bol{u}}B=0$ and $\iota_{\bol{u}}B\in H^{1}\Lambda^{1}_D(\Omega)$. By the induction equation \eqref{Bt}, $d(A_t+\iota_{\bol{u}}B)+\xi_t=0$. By $A_t+\iota_{\bol{u}}B\in H^{1}\Lambda^{1}_{D}(\Omega)$ and uniqueness of Hodge--Morrey decomposition (Lemma \ref{l: HM}), $\xi_{t} = 0$. By taking the trace of $B=dA+\xi$ at time zero, $\xi=0$ follows.
\end{proof}

\begin{thm}\label{t: thmg}
Let $\Omega$ be a compact and orientable $(2m+1)$-dimensional Riemannian manifold with smooth boundary for $m\geq 1$. The magnetic helicity \eqref{MH2} is constant for smooth solutions to the ideal MHD \eqref{iMHD} for $B_0\in dH^{1}\Lambda_D^{1}(\Omega)$ subject to the boundary conditions $\bol{u}\cdot \bol{n}=0$ and $i^{*}B=0$ on $\partial\Omega$.
\end{thm}

\begin{proof}
By Propositions \ref{p: ex} and \ref{p: ex2}, the  
magnetic field 2-form $B$ 
arising from a smooth solution of ideal MHD 
\eqref{iMHD} 
is exact and satisfies the orthogonal condition \eqref{eq: FC3}. We take an arbitrary $A$ such that $B=dA$. By the induction equation \eqref{Bt}, $d(A_t+\iota_{\bol{u}}B)=0$ and there exists $\Phi\in H^{1}\Lambda_D^{0}(\Omega)$ such that $A_t+\iota_{\bol{u}}B+d\Phi=0$. By the same argument as in the proof of Proposition \ref{p: MHC}, the magnetic helicity is conserved for all time. 
\end{proof}

\section{Generalized mean-square magnetic potential}

We show the conservation of the generalized mean-square magnetic potential~\eqref{eq: P} for symmetric solutions to the ideal MHD system~\eqref{iMHD} in \( \Omega = \Sigma \times \mathbb{R} \), thereby completing the proof of Theorem~\ref{t: thm2}. 
First, we establish the existence of a symmetric \(1\)-form \( A = A_i\,dx^i \) such that \( B = dA \). 
We then show that the symmetric component \( A_n \) evolves according to a Lie--advection equation on~\( \Sigma \).


\begin{prop}\label{p: VP}
Let $\Sigma$ be a compact and orientable $(n-1)$-dimensional Riemannian manifold with smooth boundary for $n\geq 3$ with trivial first and second cohomology groups. Let $\Omega=\Sigma \times \mathbb{R}$. Let $B$ be a closed $2$-form whose coefficients are independent of the coordinate $x^n\in\mathbb{R}$.
Then, there exists a $1$-form $A$ with coordinate expression $A=A_{i}\,dx^{i}$ such that $A_i$ is independent of $x^n$ and $B=dA$. 
\end{prop}

\begin{proof}
We express $B=p+q\wedge dx^{n}$ by the $2$-form $p=p_{ij}dx^{i}\wedge dx^{j}$ for $i,j\neq n$ and the $1$-form $q=q_idx^{i}$ for $i\neq n$. Since the coefficients of $B$ are independent of the $x^n$ variable, 
\begin{align*}
0=dB=d_{\Sigma}p+d_{\Sigma}q\wedge dx^{n},
\end{align*}
for the exterior derivative $d_{\Sigma}$ in the  coordinates $(x^1,\cdots,x^{n-1})$ on $\Sigma$. By Hodge--Morrey decomposition (Lemma \ref{l: HM}), there exist a 1-form $A_idx^{i}\in H^{1}\Lambda_{D}^{1}(\Sigma)$ and a $0$-form $A_{n}\in H^{1}\Lambda_{D}^{0}(\Sigma)$ such that $p=d_{\Sigma}(A_idx^{i})$ and $q=d_{\Sigma}A_n$. We set $A=A_{i}dx^{i}+A_ndx^{n}$ so that 
\begin{align*}
dA=d_{\Sigma}(A_idx^{i}+A_ndx^{n})=p+q\wedge dx^{n}=B.
\end{align*}
\end{proof}

\begin{proof}[Proof of Theorem \ref{t: thm2}]
For smooth solutions $(\nu,u,B,h)$ to \eqref{iMHD} in $\Omega=\Sigma\times \mathbb{R}$ such that the coefficients of $(\nu,u,B,h)$  are independent of $x^n$, we take a $1$-form $A=A_i\,dx^{i}$ such that $B=dA$ and $A_i$ are independent of $x^n$ by Proposition \ref{p: VP}. 

We show that $A_n$ is transported on $\Sigma$ by the vector field $\bol{v}=\sum_{i\neq n}u^{i}\partial_i$. 
By the induction equation  \eqref{Bt}, 
\begin{align*}
d(A_t+\iota_{\bol{u}}(dA))=0.
\end{align*}
We observe that 
\begin{align*}
\iota_{\bol{u}}dA=\sum_{i\neq n}\iota_{\bol{u}}(dA_i\wedge dx^{i})+\iota_{\bol{u}}(dA_n\wedge dx^{n})
=\sum_{i\neq n}\lr{\lr{\mf{L}_{\bol{u}}A_i} dx^{i}+u^{i}dA_i}+\lr{\mf{L}_{\bol{u}}A_n} dx^{n}+u^{n}dA_n, 
\end{align*}
and that the $dx^{n}$ component of $\iota_{\bol{u}}dA$ is $\mf{L}_{\bol{u}}A_n$. By differetiating the $dx^{i}$ and $dx^{n}$ components of $A_t+\iota_{\bol{u}}dA$, 
\begin{align*}
d_{\Sigma}\left( (A_{n})_t+\mf{L}_{\bol{u}}A_n \right)\wedge dx^{n}+d_{\Sigma}\left( \sum_{i\neq n}\lr{(A_{i})_t+\mf{L}_{\bol{u}}A_i dx^{i}+u^{i}dA_i}
+u^{n}dA_n \right)=0.
\end{align*}
Thus, $d_{\Sigma}\left( (A_{n})_t+\mf{L}_{\bol{u}}A_n \right)=0$ and hence $(A_{n})_t+\mf{L}_{\bol{u}}A_n=C$ for some function $C\lr{x^n}$. 
By the boundary conditions $i^{*}A_n=0$ and $\bol{u}\cdot \bol{n}=0$ on $\partial\Sigma$, $i^{*}\mf{L}_{\bol{u}}A_n=i^{*}\iota_{\bol{u}}dA_n=0$ on $\partial\Sigma$. Thus, $C=0$. Since  
\begin{align*}
\mf{L}_{\bol{u}}A_n=\iota_{\bol{v}+u^{n}\partial_n}dA_n=\iota_{\bol{v}}d_{\Sigma}A_n+\iota_{u^{n}\partial_n}d_{\Sigma}A_n=v^{i}\partial_iA_n, 
\end{align*}
the function $A_n$ satisfies 
\begin{align*}
(A_{n})_t+v^{i}\partial_iA_n=0.
\end{align*}
By $\nu_t+\mf{L}_{\bol{u}}\nu=0$, $\rho_t+\textrm{div}(\rho\bol{u})=0$. By $\textrm{div}\ \bol{u}=\sum_{i\neq n}\frac{1}{\sqrt{g}}\partial_i\lr{\sqrt{g}u^{i}}=\textrm{div}_{\Sigma}\bol{v}$, $\rho_t+\textrm{div}_{\Sigma}(\rho\bol{v})=0$.   
By using the continuity equation  
\begin{align*}
\left(f(A_n) \rho \right)_t+\textrm{div}_{\Sigma}\left(   f\left(A_n\right)\rho \bol{v}  \right)=0. 
\end{align*} 
For $dV=dV_{\Sigma}\wedge dx^{n}$, we set $\sigma=\rho\, dV_{\Sigma}$ so that $\nu =\sigma \wedge dx^{n}$. By using $d\iota_{\bol{v}}(f(A_n)\rho\, dV_{\Sigma})=\textrm{div}_{\Sigma} (\bol{v}f(A_n)\rho )\,dV_{\Sigma}$,  
\begin{align*}
\left(f\left(A_n \right)\sigma \right)_t+d_{\Sigma}\iota_{\bol{v}}\left(  f\left(A_n \right) \sigma \right)=0. 
\end{align*}
The conservation \eqref{eq: P} follows from Stokes's theorem and $\bol{v}\cdot \bol{n}=0$ on $\partial\Sigma$.
\end{proof}



\appendix

\section{Fundamental invariants}

We show conservation of total mass and total energy for smooth solutions to the ideal MHD \eqref{eq: IMHD} subject to the boundary conditions $\bol{u}\cdot \bol{n}=0$ and $i^*B=0$ on $\partial\Omega$ by setting $\mc{U}=\frac{1}{\rho}\int \tilde{h}\,d\rho$ and $\tilde{h}(\rho)=h+\frac{1}{2}|\bol{u}|^{2}$ for the inhomogeneous case $\rho\neq \textrm{const}$. and $\mc{U}=\mc{U}(\rho_0)$ and an arbitrary constant $\mc{U}(\rho_0)$ for the homogeneous (incompressible) case  $\rho=\rho_0$. We also show the conservation of total momentum by assuming that the mechanical pressure $P=\rho\lr{\tilde{h}-\mc{U}}$ balances with the magnetic pressure  $\abs{B}^2/4\mu_0$ on $\partial\Omega$.

\begin{thm}
Let $(\nu,u,B,h)$ be a smooth solution to the ideal MHD equations \eqref{iMHD} satisfying the boundary conditions $\bol{u}\cdot \bol{n}=0$, $i^*B=0$, and $\rho\lr{\tilde{h}-\mc{U}}+\abs{B}^2/4\mu_0=0$ on $\partial\Omega$. Then, the following quantities are conserved:

\noindent
(i) The total mass
 \eq{
 N=\int_{\Omega}\nu\label{N}
 }

 
\noindent
(ii) The total energy 
\begin{equation}
H=\displaystyle\int_{\Omega}\lr{\frac{\abs{\bol{u}}^2}{2}+\mc{U}(\rho)+\frac{\abs{B}^2}{4\mu_0\rho}}\nu  \label{H4}
\end{equation}
\\
(iii) The total $i$th momentum 
\eq{
P^i=\int_{\Omega}u^i\nu,~~~~i=1,...,n\label{Pi}
}  
\end{thm}

\begin{proof} 
(i) The conservation of \eqref{N} follows from the continuity equation $\nu_t+d\iota_{\bol{u}}\nu=0$ and the boundary condition $\bol{u}\cdot \bol{n}=0$ on $\partial \Omega$ by applying Stokes' theorem. 

(ii) The total energy \eqref{H4} 
can be written as $H=\int_{\Omega}\mc{E}$ by the energy density $n$-form
\begin{equation*}
\mc{E}=\lr{\frac{\abs{\bol{u}}^2}{2}+\mc{U}\lr{\rho}+\frac{\abs{B}^2}{4\mu_0\rho}}\nu.\label{E}
\end{equation*}
By using $(\rho \mc{U}(\rho))_{\rho}=\tilde{h}(\rho)=h+\frac{1}{2}|\bol{u}|^{2}$ and $\abs{B}^{{{2}}}=B_{ij}B^{ij}$, and $\nu_t+d\iota_{\bol{u}}\nu=0$, 
\begin{align*}
\mc{E}_t=&\lr{\frac{1}{2}\rho_t\abs{\bol{u}}^2+\rho\iota_{\p_i}\frac{\p u}{\p t}u^i+\frac{\p\lr{\rho\,\mc{U}}}{\p\rho}\rho_t+\frac{B^{ij}}{2\mu_0}\frac{\p B_{ij}}{\p t}}\,dV\\
=&-{\rm div}\lr{\rho\bol{u}}\lr{h+\abs{\bol{u}}^2}\,dV-\rho\iota_{\bol{u}}\lr{\iota_{\bol{u}}du+\frac{1}{\rho}\iota_{\bol{J}}B+dh+d\abs{\bol{u}}^2}\,dV+\frac{B^{ij}}{2\mu_0}\frac{\p B_{ij}}{\p t}\,dV\\
=&-{\rm div}\lrs{\lr{h+\abs{\bol{u}}^2}\rho\bol{u}}\,dV+\lr{\frac{B^{ij}}{2\mu_0}\frac{\p B_{ij}}{\p t}-u^jJ^iB_{ij}}\,dV.\label{Ht1}
\end{align*}
By the induction equation $B_t=-d\iota_{\bol{u}}B$, 
\begin{equation*}
\frac{\p B_{ij}}{\p t}=\nabla_j\lr{u^kB_{ki}}-\nabla_i\lr{u^kB_{kj}}.
\end{equation*}
By multiplying $B^{ij}/2\mu_0$ by the equation and summing up for $i$ and $j$,
\begin{align*}
\frac{B^{ij}}{2\mu_0}\frac{\p B_{ij}}{\p t}=\frac{B^{ij}}{\mu_0}\nabla_j\lr{u^kB_{ki}}.\label{iH}
\end{align*}
By using $\mu_0\bol{J}=-\textrm{div}\ B^{\sharp}$, 
\begin{align*}
-u^kJ^iB_{ik}=
-\frac{1}{\mu_0}u^kB_{ik}\nabla_jB^{ij}.  
\end{align*}
By substituting the above two expressions of the magnetic $2$-form into the identity of the energy density, we obtain 
\begin{align*}
\mc{E}_t=&-d\iota_{\bol{u}}\lr{h+\abs{\bol{u}}^2}\nu+\frac{1}{\mu_0}\nabla_k\lr{B^{ki}u^jB_{ij}}\,dV\\
=&-d\iota_{\bol{u}}\lr{h+\abs{\bol{u}}^2}\nu+\frac{1}{\mu_0}d\iota_{B^{ki}u^jB_{ij}\p_k}\,dV
=d\iota_{\bol{\xi}}\nu,
\end{align*}
for the vector field 
\begin{align*}
\bol{\xi}=-\lr{h+\abs{\bol{u}}^2}\bol{u}+\frac{1}{\mu_0\rho}B^{ki}u^jB_{ij}\p_k.
\end{align*}
By the boundary conditions $\bol{u}\cdot\bol{n}=0$ and $i^{\ast}B=0$ on $\partial \Omega$, 
\begin{align*}
\rho\,\bol{\xi}\cdot\bol{n}=&-\rho\lr{h+\abs{\bol{u}}^2}\bol{u}\cdot\bol{n}-\frac{1}{\mu_0}g^{\sharp}\lr{\iota_{\bol{n}}B,\iota_{\bol{u}}B}=0.
\end{align*}
The conservation of \eqref{H4} follows from Stokes' theorem.  

(iii) By the closure condition $dB=0$,  
\begin{equation*} \nabla_kB_{ij}+\nabla_iB_{jk}+\nabla_{j}B_{ki}=0.
\end{equation*}
Observe that 
\begin{align*}
\nabla_k\lr{ B^{kj}B_{j\ell}g^{i\ell}+\frac{1}{4}g^{ki}\abs{B}^2}=&-\mu_0J^jB_{j\ell}g^{i\ell}+g^{i\ell} B^{kj}\nabla_k B_{j\ell}+\frac{1}{4}\nabla_k\lr{g^{ki}\abs{B}^2}\\
=&\mu_0f^{i}+\frac{1}{2}g^{i\ell}B^{kj}\lr{\nabla_kB_{j\ell}+\nabla_jB_{\ell k}}+\frac{1}{4}\nabla_k\lr{g^{ki}\abs{B}^2}\\
=&\mu_0f^i-\frac{1}{2}g^{i\ell}B^{kj}\nabla_{\ell}B_{kj}+\frac{1}{4}\nabla_k\lr{g^{ki}\abs{B}^2}\\
=&\mu_0f^i,
\end{align*}
where $f^i$ denotes the $i$th component of the Lorentz force. 
By using the identity
\begin{equation*}
{\rm div}\lr{\bol{u}}=\frac{1}{\sqrt{g}}\p_i\lr{\sqrt{g}u^i}=\nabla_i u^i,
\end{equation*}
it follows that 
\begin{equation*}
\lr{u^i\nu}_t=-\nabla_k\lrs{\rho u^ku^i+
\rho\lr{
\tilde{h}-\mc{U}
-\frac{\abs{B}^2}{4\mu_0\rho}}g^{ki}-\frac{1}{\mu_0}B^{kj}B_{j\ell}g^{i\ell}}\,dV
=-d\iota_{\bol{v}^i}\nu,
.\label{dPdtmom}
\end{equation*}
for
\begin{align*}
\bol{v}^i=u^i\bol{u}+
\lr{
\tilde{h}-\mc{U}
-\frac{\abs{B}^2}{4\mu_0\rho}}g^{ik}\p_k-\frac{1}{\mu_0\rho}B^{kj}B_{j\ell}g^{\ell i}\p_k.
\end{align*}
By taking the inner product with $\bol{n}$, 
\begin{align*}
\rho\bol{v}^i\cdot\bol{n}=\rho u^i\bol{u}\cdot\bol{n}+\rho\lr{\tilde{h}-\mc{U}-\frac{\abs{B}^2}{4\mu_0\rho}}n^i+\frac{1}{\mu_0} g^{\sharp}\lr{\iota_{\bol{n}}B,\iota_{\p^i}B},\label{vin}
\end{align*}
where $n^i=\iota_{\bol{n}}dx^i$ and  $\p^i=g^{ij}\p_j$. 
Since $i^*B=0$ on $\partial\Omega$, we may write $B=\bol{n}^{\flat}\w \eta$ for some tangential $1$-form $\eta$ such that $\iota_{\bol{n}}\eta=0$. Then, $g^{\sharp}\lr{\iota_{\bol{n}}B,\iota_{\p^i}B}=n^i\abs{\eta}^2=n^i\abs{B}^2/2$ and  
\begin{align*}
\rho\bol{v}^i\cdot\bol{n}=\rho u^i\bol{u}\cdot\bol{n}+\rho\lr{\tilde{h}-\mc{U}+\frac{\abs{B}^2}{4\mu_0\rho}}n^i=0.
\end{align*}
The conservation of \eqref{Pi} follows from Stokes' theorem.
\end{proof}

\bibliographystyle{alphaurl}

\begin{thebibliography}{MMMT86}

\bibitem[AK22]{ArnoldKh}
V.~I. Arnold and B.~A. Khesin.
\newblock {\em Topological Methods in Hydrodynamics}.
\newblock Springer, 2 edition, 2022.
\newblock \href {https://doi.org/10.1007/978-3-030-74278-2} {\path{doi:10.1007/978-3-030-74278-2}}.

\bibitem[Arn66]{A66}
V.~I. Arnold.
\newblock Sur la g\'{e}om\'{e}trie diff\'{e}rentielle des groupes de {L}ie de dimension infinie et ses applications \`a l'hydrodynamique des fluides parfaits.
\newblock {\em Ann. Inst. Fourier (Grenoble)}, 16(fasc. 1):319--361, (1966).
\newblock \href {https://doi.org/10.5802/AIF.233} {\path{doi:10.5802/AIF.233}}.

\bibitem[BFV22]{BFV21}
R.~Beekie, S.~Friedlander, and V.~Vicol.
\newblock On {M}offatt's magnetic relaxation equations.
\newblock {\em Comm. Math. Phys.}, 390:1311--1339, (2022).
\newblock \href {https://doi.org/10.1007/s00220-021-04289-3} {\path{doi:10.1007/s00220-021-04289-3}}.

\bibitem[BG80]{BG}
M.~Bevir and J.~W. Gray.
\newblock Relaxation, flux consumption and quasi steady state pinches.
\newblock In {\em Proc. of RFP Theory Workshop}, volume LA-8944-C, pages 176--180, 1980.

\bibitem[Bis03]{Bis03}
D.~Biskamp.
\newblock {\em Magnetohydrodynamic Turbulence}.
\newblock Cambridge University Press, 2003.
\newblock \href {https://doi.org/10.1017/CBO9780511535222} {\path{doi:10.1017/CBO9780511535222}}.

\bibitem[BKS25]{BKS23}
H.~Bae, H.~Kwon, and J.~Shin.
\newblock {G}lobal solutions to {S}tokes-{M}agneto equations with fractional dissipations.
\newblock {\em Partial Differ. Equ. Appl.}, 6(4), (2025).
\newblock \href {https://doi.org/10.1007/s42985-025-00341-2} {\path{doi:10.1007/s42985-025-00341-2}}.

\bibitem[BS03]{Bau03}
T.~W. Baumgarte and S.~L. Shapiro.
\newblock General relativistic magnetohydrodynamics for the numerical construction of dynamical spacetimes.
\newblock {\em The Astrophysical Journal}, 585:921--929, (2003).
\newblock \href {https://doi.org/doi.org/10.1086/346103} {\path{doi:doi.org/10.1086/346103}}.

\bibitem[BV22]{BV}
J.~Bedrossian and V.~Vicol.
\newblock {\em The Mathematical Analysis of the Incompressible Euler and Navier--Stokes Equations}.
\newblock American Mathematical Society, 2022.
\newblock \href {https://doi.org/10.1090/gsm/225} {\path{doi:10.1090/gsm/225}}.

\bibitem[CDG21a]{CDG21b}
P.~Constantin, T.~D. Drivas, and D.~Ginsberg.
\newblock Flexibility and rigidity in steady fluid motion.
\newblock {\em Commun. Math. Phys.}, 385:521--563, (2021).
\newblock \href {https://doi.org/10.1007/s00220-021-04048-4} {\path{doi:10.1007/s00220-021-04048-4}}.

\bibitem[CDG21b]{CDG21}
P.~Constantin, T.~D. Drivas, and D.~Ginsberg.
\newblock On quasisymmetric plasma equilibria sustained by small force.
\newblock {\em J. Plasma Phys.}, 87:905870111, (2021).
\newblock \href {https://doi.org/10.1017/S0022377820001610} {\path{doi:10.1017/S0022377820001610}}.

\bibitem[CDP25]{CDP}
R.~Cardona, N.~Duignan, and D.~Perrella.
\newblock {A}symmetry of {M}{H}{D} {E}quilibria for {G}eneric {A}dapted {M}etrics.
\newblock {\em Arch. Rational Mech. Anal.}, 249(1), (2025).
\newblock \href {https://doi.org/10.1007/s00205-024-02075-8} {\path{doi:10.1007/s00205-024-02075-8}}.

\bibitem[CP23]{CP22}
P.~Constantin and F.~Pasqualotto.
\newblock {M}agnetic {R}elaxation of a {V}oigt--{M}{H}{D} {S}ystem.
\newblock {\em Commun. Math. Phys.}, 402:1931--1952, (2023).
\newblock \href {https://doi.org/10.1007/s00220-023-04770-1} {\path{doi:10.1007/s00220-023-04770-1}}.

\bibitem[Dav01]{Davidson}
P.~A. Davidson.
\newblock {\em An introduction to magnetohydrodynamics}.
\newblock Cambridge Texts in Applied Mathematics. Cambridge University Press, Cambridge, 2001.
\newblock \href {https://doi.org/10.1017/CBO9780511626333} {\path{doi:10.1017/CBO9780511626333}}.

\bibitem[DE23]{DE}
T.~D. Drivas and T.~M. Elgindi.
\newblock Singularity formation in the incompressible {E}uler equation in finite and infinite time.
\newblock {\em EMS Surv. Math. Sci.}, 10(1):1--100, (2023).
\newblock \href {https://doi.org/10.4171/emss/66} {\path{doi:10.4171/emss/66}}.

\bibitem[Dez83]{Dez}
A.~A. Dezin.
\newblock Invariant forms and some structure properties of the {E}uler equations of hydrodynamics.
\newblock {\em Zeit. Anal. Anwend. (in Russian)}, 2:401--409, (1983).

\bibitem[DMP15]{DAvignon15}
E.~D'Avignon, P.~J. Morrison, and F.~Pegoraro.
\newblock Action principle for relativistic magnetohydrodynamics.
\newblock {\em Physical Review D}, 91:084050, (2015).
\newblock \href {https://doi.org/10.1103/PhysRevD.91.084050} {\path{doi:10.1103/PhysRevD.91.084050}}.

\bibitem[Fit14]{Fitz}
R.~Fitzpatrick.
\newblock {\em Plasma Physics: An Introduction}.
\newblock CRC Press, 2014.
\newblock \href {https://doi.org/10.1201/b17263} {\path{doi:10.1201/b17263}}.

\bibitem[FLS22]{Faraco}
D.~Faraco, S.~Lindberg, and L.~Sz{\'e}kelyhidi.
\newblock Rigorous results on conserved and dissipated quantities in ideal mhd turbulence.
\newblock {\em Geophysical \& Astrophysical Fluid Dynamics}, 116(4):237--260, (2022).
\newblock \href {https://doi.org/10.1080/03091929.2022.2060964} {\path{doi:10.1080/03091929.2022.2060964}}.

\bibitem[Fra79]{Frankel79}
T.~Frankel.
\newblock {\em Gravitational Curvature}.
\newblock Dover Publications, 1979.

\bibitem[Fra12]{TF}
T.~Frankel.
\newblock {\em The Geometry of Physics}.
\newblock Cambridge University Press, 3 edition, 2012.
\newblock \href {https://doi.org/10.1017/CBO9781139061377} {\path{doi:10.1017/CBO9781139061377}}.

\bibitem[Fre14]{Freid}
J.~P. Freidberg.
\newblock {\em {I}deal {M}{H}{D}}.
\newblock Cambridge University Press, 2014.
\newblock \href {https://doi.org/10.1017/CBO9780511795046} {\path{doi:10.1017/CBO9780511795046}}.

\bibitem[Ghr01]{Gh01}
R.~Ghrist.
\newblock Steady nonintegrable high-dimensional fluids.
\newblock {\em Lett. Math. Phys.}, 55(3):193--204, (2001).
\newblock \href {https://doi.org/10.1023/A:1010936025007} {\path{doi:10.1023/A:1010936025007}}.

\bibitem[GK92]{GK1}
V.~L. Ginzburg and B.~A. Khesin.
\newblock Topology of steady fluid flows.
\newblock In eds. H.K. Moffatt~et al., editor, {\em Topological aspects of the dynamics of fluids and plasmas}, pages 265--272. Kluwer Acad. Publ., Dordrecht, (1992).
\newblock \href {https://doi.org/10.1007/978-94-017-3550-6} {\path{doi:10.1007/978-94-017-3550-6}}.

\bibitem[GK94]{GK2}
V.~L. Ginzburg and B.~A. Khesin.
\newblock Steady fluid flows and symplectic geometry.
\newblock {\em J. Geometry and Physics}, 14(2):195--210, (1994).
\newblock \href {https://doi.org/10.1016/0393-0440(94)90006-X} {\path{doi:10.1016/0393-0440(94)90006-X}}.

\bibitem[Gra67]{Grad67}
H.~Grad.
\newblock Toroidal containment of a plasma.
\newblock {\em The Physics of Fluids}, 10:137--154, (1967).

\bibitem[Gra85]{Grad85}
H.~Grad.
\newblock Theory and applications of the nonexistence of simple toroidal plasma equilibrium.
\newblock {\em Int. J. Fusion Energy}, 3:33--46, (1985).

\bibitem[GV20]{GV}
A.~D. Gilbert and J.~Vanneste.
\newblock A geometric look at {MHD} and the {B}raginsky dynamo.
\newblock {\em Geophys. Astrophys. Fluid Dyn.}, 115(4):436--471, (2020).
\newblock \href {https://doi.org/10.1080/03091929.2020.1839896} {\path{doi:10.1080/03091929.2020.1839896}}.

\bibitem[GV23]{GV2}
A.~D. Gilbert and J.~Vanneste.
\newblock A geometric look at momentum flux and stress in fluid mechanics.
\newblock {\em Journal of Nonlinear Science}, 33(2):31, (2023).
\newblock \href {https://doi.org/10.1007/s00332-023-09887-0} {\path{doi:10.1007/s00332-023-09887-0}}.

\bibitem[Has85]{Hasegawa85}
A.~Hasegawa.
\newblock Self-organization processes in continuous media.
\newblock {\em Advances in Physics}, 34:1--42, (1985).
\newblock \href {https://doi.org/10.1080/00018738500101721} {\path{doi:10.1080/00018738500101721}}.

\bibitem[HHBB25]{Bhatt}
Y.-M. Huang, J.~K.~J. Hew, A.~Brown, and A.~Bhattacharjee.
\newblock Computation of magnetohydrodynamic equilibria with voigt regularization.
\newblock {\em Physics of Plasmas}, 32(6):062507, (2025).
\newblock \href {https://doi.org/10.1063/5.0267510} {\path{doi:10.1063/5.0267510}}.

\bibitem[HHS24]{HHS}
D.~D. Holm, R.~Hu, and O.~D. Street.
\newblock Deterministic and stochastic geometric mechanics for hall magnetohydrodynamics.
\newblock {\em Proceedings of the Royal Society A: Mathematical, Physical and Engineering Sciences}, 480:2300, (2024).
\newblock \href {https://doi.org/10.1098/rspa.2024.0267} {\path{doi:10.1098/rspa.2024.0267}}.

\bibitem[HHS25]{HHS2}
D.~D. Holm, R.~Hu, and O.~D. Street.
\newblock A variational perspective on plasma dynamics in thin domains.
\newblock {\em Nonlinearity}, 38(9), (2025).
\newblock \href {https://doi.org/10.1088/1361-6544/adffd8} {\path{doi:10.1088/1361-6544/adffd8}}.

\bibitem[HMR98]{Holm98}
D.~D. Holm, J.~E. Marsden, and T.~S. Ratiu.
\newblock The {E}uler--{P}oincar{\'e} {E}quations and {S}emidirect {P}roducts with {A}pplications to {C}ontinuum {T}heories.
\newblock {\em Advances in Mathematics}, 137(1):1--81, (1998).
\newblock \href {https://doi.org/10.1006/aima.1998.1721} {\path{doi:10.1006/aima.1998.1721}}.

\bibitem[IK17]{IK17}
A.~Izosimov and B.~Khesin.
\newblock Classification of casimirs in 2d hydrodynamics.
\newblock {\em Moscow Mathematical Journal}, 17(4):699--716, (2017).
\newblock \href {https://doi.org/10.17323/1609-4514-2017-17-4-699-716} {\path{doi:10.17323/1609-4514-2017-17-4-699-716}}.

\bibitem[KC89]{KC}
B.~A. Khesin and Yu.~V. Chekanov.
\newblock Invariants of the euler equation for ideal or barotropic hydrodynamics and superconductivity in d dimensions.
\newblock {\em Physica D}, 40(1):119--131, (1989).
\newblock \href {https://doi.org/10.1016/0167-2789(89)90030-4} {\path{doi:10.1016/0167-2789(89)90030-4}}.

\bibitem[KMeS23]{KMS23}
B.~Khesin, G.~Misio\l~ek, and A.~Shnirelman.
\newblock Geometric hydrodynamics in open problems.
\newblock {\em Arch. Ration. Mech. Anal.}, 247(2):Paper No. 15, 43, (2023).
\newblock \href {https://doi.org/10.1007/s00205-023-01848-x} {\path{doi:10.1007/s00205-023-01848-x}}.

\bibitem[KMM17]{Kaw17}
Y.~Kawazura, G.~Miloshevich, and P.~J. Morrison.
\newblock Action principles for relativistic extended magnetohydrodynamics: A unified theory of magnetofluid models.
\newblock {\em Physics of Plasmas}, 24:022103, (2017).
\newblock \href {https://doi.org/doi.org/10.1063/1.4975013} {\path{doi:doi.org/10.1063/1.4975013}}.

\bibitem[KPSY21]{Khesin21}
B.~Khesin, D.~Peralta-Salas, and C.~Yang.
\newblock A basis of {C}asimirs in 3{D} magnetohydrodynamics.
\newblock {\em Int. Math. Res. Not. IMRN}, (18):13645--13660, (2021).
\newblock \href {https://doi.org/10.1093/imrn/rnz393} {\path{doi:10.1093/imrn/rnz393}}.

\bibitem[Liz22]{Liz}
F.~T.~de Lizaur.
\newblock Chaos in the incompressible {E}uler equation on manifolds of high dimension.
\newblock {\em Invent. Math.}, 228(2):687--715, (2022).
\newblock \href {https://doi.org/10.1007/s00222-021-01089-3} {\path{doi:10.1007/s00222-021-01089-3}}.

\bibitem[LMM16]{Mo}
M.~Lingam, G.~Miloshevich, and P.~J. Morrison.
\newblock Concomitant hamiltonian and topological structures of extended magnetohydrodynamics.
\newblock {\em Physics Letters A}, 380(31):2400--2406, (2016).
\newblock \href {https://doi.org/10.1016/j.physleta.2016.05.024} {\path{doi:10.1016/j.physleta.2016.05.024}}.

\bibitem[MMMT86]{Marsden86}
J.~E. Marsden, R.~Montgomery, P.~J. Morrison, and W.~B. Thompson.
\newblock Covariant poisson brackets for classical fields.
\newblock {\em Annals of Physics}, 169:29--47, (1986).
\newblock \href {https://doi.org/10.1016/0003-4916(86)90157-0} {\path{doi:10.1016/0003-4916(86)90157-0}}.

\bibitem[Mof69]{Moffatt69}
H.~K. Moffatt.
\newblock The degree of knottedness of tangled vortex lines.
\newblock {\em J. Fluid Mech.}, 35(1):117--129, (1969).
\newblock \href {https://doi.org/10.1017/S0022112069000991} {\path{doi:10.1017/S0022112069000991}}.

\bibitem[Mof85]{Moffatt85}
H.~K. Moffatt.
\newblock Magnetostatic equilibria and analogous {E}uler flows of arbitrarily complex topology. {I}. {F}undamentals.
\newblock {\em J. Fluid Mech.}, 159:359--378, (1985).
\newblock \href {https://doi.org/10.1017/S0022112085003251} {\path{doi:10.1017/S0022112085003251}}.

\bibitem[Mof21]{Moffatt21}
H.~K. Moffatt.
\newblock Some topological aspects of fluid dynamics.
\newblock {\em J. Fluid Mech.}, 914(P1):1--56, 2021.
\newblock \href {https://doi.org/10.1017/jfm.2020.230} {\path{doi:10.1017/jfm.2020.230}}.

\bibitem[MRW84]{MRW}
J.~E. Marsden, T.~Ra{\c t}iu, and A.~Weinstein.
\newblock Semidirect products and reduction in mechanics.
\newblock {\em Trans. Amer. Math. Soc.}, 281:147--177, (1984).

\bibitem[MV19]{TV}
D.~MacTaggart and A.~Valli.
\newblock Magnetic helicity in multiply connected domains.
\newblock {\em Journal of Plasma Physics}, 85:775850501, (2019).
\newblock \href {https://doi.org/10.1017/S0022377819000576} {\path{doi:10.1017/S0022377819000576}}.

\bibitem[MW83]{MW}
J.~Marsden and A.~Weinstein.
\newblock Coadjoint orbits, vortices, and clebsch variables for incompressible fluids.
\newblock {\em Physica D}, 7(1-3):305--323, (1983).
\newblock \href {https://doi.org/10.1016/0167-2789(83)90134-3} {\path{doi:10.1016/0167-2789(83)90134-3}}.

\bibitem[New61]{New}
W~A Newcomb.
\newblock Lagrangian and hamiltonian methods in magnetohydrodynamics.
\newblock \href{https://www.osti.gov/biblio/5034470}{https://www.osti.gov/biblio/5034470}, 06 1961.

\bibitem[OKC88]{OKC}
V.~Yu. Ovsienko, B.~A. Khesin, and Yu.~V. Chekanov.
\newblock Integrals of the euler equations in multidimensional hydrodynamics and superconductivity.
\newblock {\em (in Rus- sian) Diff. Geom., Lie Groups, and Mechanics, Zap. Sem. LOMI 172; English transl.: J. of Sov. Math. 59:5 (1992), 1096--1102}, pages 105--113, (1988).
\newblock \href {https://doi.org/10.1007/BF01480692} {\path{doi:10.1007/BF01480692}}.

\bibitem[PNP21]{Pfeff21}
D.~Pfefferl{\'e}, L.~Noakes, and D.~Perrella.
\newblock Gauge freedom in magnetostatics and the effect on helicity in toroidal volumes.
\newblock {\em Journal of Mathematical Physics}, 62:093505, (2021).
\newblock \href {https://doi.org/10.1063/5.0038226} {\path{doi:10.1063/5.0038226}}.

\bibitem[Sch95]{Schwarz}
G.~Schwarz.
\newblock {\em Hodge Decomposition: A Method for Solving Boundary Value Problems}.
\newblock Lecture Notes in Mathematics. Springer, 1995.
\newblock \href {https://doi.org/10.1007/BFb0095978} {\path{doi:10.1007/BFb0095978}}.

\bibitem[Ser84]{Ser}
D.~Serre.
\newblock Invariants et d{\'e}g{\'e}n{\'e}rescence symplectique de l'{\'e}quation d'euler des fluides parfaits incompressibles.
\newblock {\em C.R. Acad. Sci. Paris, S\'er. A}, 298:349--352, (1984).

\bibitem[Ser18]{Serre18}
D.~Serre.
\newblock Helicity and other conservation laws in perfect fluid motion.
\newblock {\em Comptes Rendus M{\'e}canique}, 346(3):175--183, (2018).
\newblock The legacy of Jean-Jacques Moreau in mechanics / L'h{\'e}ritage de Jean-Jacques Moreau en m{\'e}canique.
\newblock \href {https://doi.org/10.1016/j.crme.2017.12.001} {\path{doi:10.1016/j.crme.2017.12.001}}.

\bibitem[Tan]{JINTAN}
J.~Tan.
\newblock Weak solutions of moffatt's magnetic relaxation equations.
\newblock \href{https://arxiv.org/abs/2311.18407}{arXiv:2311.18407}.

\bibitem[Tao18]{Tao18}
T.~Tao.
\newblock On the universality of the incompressible euler equation on compact manifolds.
\newblock {\em Discrete Contin. Dyn. Syst.}, 38(3):1553--1565, (2018).
\newblock \href {https://doi.org/10.3934/dcds.2018064} {\path{doi:10.3934/dcds.2018064}}.

\bibitem[Tao20]{Tao20}
T.~Tao.
\newblock On the universality of the incompressible {E}uler equation on compact manifolds, {II}. {N}on-rigidity of {E}uler flows.
\newblock {\em Pure Appl. Funct. Anal.}, 5(6):1425--1443, (2020).

\bibitem[Tay74]{Taylor74}
J.~B. Taylor.
\newblock Relaxation of toroidal plasma and generation of reverse magnetic fields.
\newblock {\em Physical Review Letters}, 33:1139--1141, 1974.
\newblock \href {https://doi.org/10.1103/PhysRevLett.33.1139} {\path{doi:10.1103/PhysRevLett.33.1139}}.

\bibitem[Tay86]{Taylor86}
J.~B. Taylor.
\newblock Relaxation and magnetic reconnection in plasmas.
\newblock {\em Reviews of Modern Physics}, 58(3), 1986.
\newblock \href {https://doi.org/10.1103/RevModPhys.58.741} {\path{doi:10.1103/RevModPhys.58.741}}.

\bibitem[VMI96]{Moffatt96}
V.~A. Vladimirov, H.~K. Moffatt, and K.~I. Ilin.
\newblock On general transformations and variational principles for the magnetohydrodynamics of ideal fluids. part {II}. {S}tability criteria for two-dimensional flows.
\newblock {\em J. Fluid Mech.}, 329:187--205, (1996).
\newblock \href {https://doi.org/10.1017/S0022112096008890} {\path{doi:10.1017/S0022112096008890}}.

\bibitem[VMI97]{Moffatt97}
V.~A. Vladimirov, H.~K. Moffatt, and K.~I. Ilin.
\newblock {O}n general transformations and variational principles for the magnetohydrodynamics of ideal fluids. {P}art {III}. {S}tability criteria for axisymmetric flows.
\newblock {\em J. Plasma Phys.}, 57:89--120, (1997).
\newblock \href {https://doi.org/10.1017/S0022377896005272} {\path{doi:10.1017/S0022377896005272}}.

\bibitem[Wei72]{Wein72}
S.~Weinberg.
\newblock {\em Gravitation and Cosmology: Principles and Applications of the General Theory of Relativity}.
\newblock John Wiley \& Sons Inc., New York, 1972.

\bibitem[YM02]{Yoshida02}
Z.~Yoshida and S.~M. Mahajan.
\newblock Variational principles and self-organization in two-fluid plasmas.
\newblock {\em Physical Review Letters}, 88(9), (2002).
\newblock \href {https://doi.org/10.1103/PhysRevLett.88.095001} {\path{doi:10.1103/PhysRevLett.88.095001}}.

\bibitem[YM17]{ZY17}
Z.~Yoshida and P.~J. Morrison.
\newblock Epi-two-dimensional fluid flow: A new topological paradigm for dimensionality.
\newblock {\em Phys. Rev. Lett.}, 119:244501, Dec (2017).
\newblock \href {https://doi.org/10.1103/PhysRevLett.119.244501} {\path{doi:10.1103/PhysRevLett.119.244501}}.

\end{thebibliography}

\end{document}